\documentclass[sigconf,authorversion,nonacm]{acmart}
\usepackage{booktabs} % For formal tables

\usepackage{balance}
\usepackage{epsfig}
\usepackage{epstopdf}

\usepackage{amsmath}
\usepackage{mathtools}

\usepackage{algorithm}
\usepackage{algpseudocode}
\newcommand{\algrule}[1][.2pt]{\par\vskip.5\baselineskip\hrule height #1\par\vskip.5\baselineskip}

\usepackage{bm}

\usepackage{multirow}

\usepackage{tabularx}
\usepackage{amsfonts}

\usepackage{microtype}

\usepackage{textcomp}

\usepackage{url}
\usepackage{url}

\DeclareMathOperator*{\argmax}{arg\,max}
\DeclareMathOperator*{\argmin}{arg\,min}
\DeclareMathOperator{\sign}{sgn}

\newcommand{\approxD}{\xrightarrow{D}}
\newcommand{\Osymbol}{{\mathcal O}}
\newcommand{\BO}[1]{\Osymbol\left(#1\right)}
\newcommand{\TilO}[1]{\tilde{\Osymbol}\left(#1\right)}

\newcommand{\E}[1]{\textrm{\bf E}\left[#1\right]}
\renewcommand{\Pr}[1]{\textrm{\bf Pr}\left[#1\right]}
\newcommand{\Var}[1]{\textrm{\bf Var}\left[#1\right]}

\newcommand{\Rd} {\mathbb{R}^d}

\newcommand{\mX} {\mathbb{X}}
\newcommand{\bX} {\mathbf{X}}
\newcommand{\bR} {\mathbf{R}}

\newcommand{\bH} {\mathbf{H}}
\newcommand{\bD} {\mathbf{D}}

\newcommand{\bp} {\mathbf p}
\newcommand{\bq} {\mathbf q}
\newcommand{\br} {\mathbf r}
\newcommand{\bx} {\mathbf x}

\newcommand{\by} {\mathbf y}

\newcommand{\dotxr}{\bx^\top \br}
\newcommand{\dotxq}{\bx^\top \bq}
\newcommand{\dotyq}{\by^\top \bq}
\newcommand{\thetaxq}{\theta_{\bx \bq}}

\newcommand{\rhox}{\rho_{\bx}}
\newcommand{\rhoy}{\rho_{\by}}

\newcommand{\osUr}{U_{(r)}}

\newcommand{\cosEr}{\epsilon_{[r]}}
\newcommand{\cosVr}{V_{[r]}}
\newcommand{\cosVone}{V_{[1]}}
\newcommand{\cosVd}{V_{[D]}}

\newcommand{\osQone}{Q_{(1)}}
\newcommand{\osQr}{Q_{(r)}}
\newcommand{\osQd}{Q_{(D)}}

\newcommand{\cosYr}{Y_{[r]}}
\newcommand{\cosYone}{Y_{[1]}}

\newcommand{\cosXr}{X_{[r]}}
\newcommand{\cosXone}{X_{[1]}}
\newcommand{\cosXd}{X_{[D]}}
\newcommand{\rhoone}{\rho_{[1]}}
\newcommand{\sigmaone}{\sigma_{[1]}}

\newcommand{\Zcos}{Z_{ceo}}
\newcommand{\Zlsh}{Z_{lsh}}

\usepackage{xcolor}

\AtBeginDocument{%
	\providecommand\BibTeX{{%
			\normalfont B\kern-0.5em{\scshape i\kern-0.25em b}\kern-0.8em\TeX}}}

\begin{document}

\title{Sublinear Maximum Inner Product Search using \\ Concomitants of Extreme Order Statistics
}

\author{Ninh Pham}
\email{ninh.pham@auckland.ac.nz}
\orcid{0000-0001-5768-9900}
\affiliation{%
	\department{School of Computer Science}
	\institution{University of Auckland}
%	\streetaddress{P.O. Box 1212}
	\city{New Zealand}
%	\state{Ohio}
%	\postcode{43017-6221}
}

%\author{Ninh Pham}
%%\authornote{The secretary disavows any knowledge of this author's actions.}
%\affiliation{%
%	\institution{School of Computer Science \\
%		University of Auckland}
%	%\streetaddress{P.O. Box 1212}
%	\city{Auckland}
%	\state{New Zealand}
%	%\postcode{43017-6221}
%}
%\email{ninh.pham@auckland.ac.nz}

%\maketitle

\begin{abstract}

We propose a novel dimensionality reduction method for maximum inner product search (MIPS), named \emph{CEOs}, based on the theory of concomitants of extreme order statistics.
Utilizing the asymptotic behavior of these concomitants, we show that a few dimensions associated with the extreme values of the query signature are enough to estimate inner products.
Since CEOs only uses the sign of a small subset of the query signature for estimation, we can precompute all inner product estimators accurately before querying.
These properties yield a \emph{sublinear} MIPS algorithm with an exponential indexing space complexity. 
We show that our exponential space is \emph{optimal} for the $(1 + \epsilon)$-approximate MIPS in unit sphere.
The search recall of CEOs can be theoretically guaranteed under a mild condition.

To deal with the exponential space complexity, we propose two practical variants, including \emph{sCEOs-TA} and \emph{coCEOs}, that use linear space for solving MIPS.
\emph{sCEOs-TA} exploits the threshold algorithm (TA) and provides superior search recalls to competitive MIPS solvers.
\emph{coCEOs} is a data and dimension co-reduction technique and outperforms \emph{sCEOs-TA} on high recall requirements.
Empirically, they are very simple to implement and achieve at least 100x speedup compared to the bruteforce search while returning top-10 MIPS with accuracy at least 90\% on many large-scale data sets.
\end{abstract}

\maketitle

\section{Introduction}

%Nearest neighbor search (NNS) and its application.
Maximum inner product search (MIPS) is the task of, given a point set $\mX \subset \Rd$ of size $n$ and a query point $\bq \in \Rd$, finding the point $\bp \in \mX$ such that, $$\bp = \argmax_{\bx \in \mX}{\bx^\top \bq} \enspace .$$
MIPS and its variant top-$k$ MIPS, which finds the top-$k$ largest inner product points with a query, are central tasks in many real-world big data applications, for example recommender systems~\cite{KorenIEEE09,Linden03}, similarity search in high dimensions~\cite{Glove,Sundaram13}, multi-class learning~\cite{Dean13,ImageNet}, and neural network~\cite{Covington16,Spring17}.

%These applications arise in large-scale high-dimensional domains and require fast query response.
Modern collaborative filtering based recommender systems, e.g. Xbox or Netflix, often deal with very large-scale data sets and require fast response~\cite{Xbox,Netflix}.
Such recommender systems present users as a query set and items as a data set.
A large inner product value between the user and item vectors indicates that the item is relevant to the user preferences.
When the context has been used~\cite{Adomavicius15}, the learning (i.e. matrix factorization) phase cannot be done entirely offline~\cite{Xbox,YahooMusic}. 
In other words, the items vectors are also computed online and hence a high cost of constructing the index structure for MIPS will significantly degrade the system performance.

Many MIPS applications arise on streaming data where both query and data points come with a rapid rate.
For example, Twitter has to search with 400 million new tweets per day~\cite{Sundaram13}.
MIPS can also be used to reduce the computational cost of training and testing deep networks~\cite{Spring17}.
However, the very large index storage with high latency of updates will deteriorate the overall performance.

Motivated by the computational bottleneck of many MIPS applications in big data, this work addresses the following problem:

\medskip

{\em If we build a data structure for $\mX$ in $\TilO{dn}$\footnote{Polylogarithmic factors, e.g. $\log{d}\log{n}$ is absorbed in the $\tilde{\Osymbol}$-notation.} time and space, can we have a fast MIPS solver that returns the best search recall?}
%by accessing at most $b = \eta n$ data points, where $\eta$ is a small constant, e.g. 1\%.?}

\medskip

Our main finding is that existing solutions, while highly efficient in general, cannot achieve effective search recall given that the index construction requires $\TilO{dn}$ time and space complexity. 
We present novel solutions based on the concomitants of extreme order statistics.
Our algorithms are very simple to implement (few lines of Python codes), run significantly faster, and yield higher search recall than competitive MIPS solvers.

Because Gaussian random projection is a building block of our approach, and because locality-sensitive hashing is our primary alternative for comparison, we review both briefly.

\subsection{Gaussian random projections}

Random projections (RP) refer to the technique of projecting data points in $d$-dimensional spaces onto random $l$-dimensional spaces ($l < d$) via a random matrix $\bR \in \mathbb{R}^{l \times d}$.
In the reduced $l$-dimensional space, the key data properties, e.g. pairwise Euclidean distances and inner product values, are preserved with high probability.
Therefore, we can achieve a high quality approximation answer for MIPS in the reduced dimensional space.

%Therefore MIPS can significantly reduced.This can reduce the similarity search from
Presenting the point set $\mX$ as a matrix $\bX \in \mathbb{R}^{d \times n}$, we generate a Gaussian random matrix $\bR \in \mathbb{R}^{l \times d}$ whose elements are randomly sampled from the standard normal distribution $N(0, 1)$.
The signature (i.e. projected representation) of $\bx \in \bX$ is computed by $\bR \bx / \sqrt{l}$.
Since we study MIPS, we only state the relative distortion bound of inner product values in the reduced space as follows:
%which takes $O(ndk)$. Arising from the Johnson-Lindenstrauss lemma \cite{JL_RP}, the (much smaller) matrix $B$ can preserve all pairwise Euclidean distances of $A$ within an arbitrarily small factor. 
%
\begin{lemma}\label{lm:JL}
Let $\thetaxq$ be the angle between the vectors $\bx, \bq \in \Rd$. 
Given $0 < \epsilon < 1$, we have the following:
\begin{align*}
%\Pr{\left(\bR \bx\right)^\top \bR \bq > \left(1 + \epsilon \right) \dotxq} &< \exp\left(-\frac{l}{8}\epsilon^2 \cos^2{(\thetaxq)} \right) \, , \\
%%
%\Pr{\left(\bR \bx\right)^\top \bR \bq < \left(1 - \epsilon \right) \dotxq} &< \exp\left(-\frac{l}{8}\epsilon^2 \cos^2{(\thetaxq)} \right)
%
\Pr{|\left(\bR \bx\right)^\top \bR \bq - \dotxq| \geq  \epsilon \dotxq} &\leq 2 \exp\left(-\frac{l}{8}\epsilon^2 \cos^2{(\thetaxq)} \right) \, .
\end{align*}
\end{lemma}

The proof of Lemma~\ref{lm:JL} can be found in~\cite{Kaban15}.
The constant of this bound on the inner product  matches the Johnson-Lindenstrauss lemma bounds on the Euclidean distance~\cite{Dasgupta03,JL}, which has been proven to be optimal for linear dimensionality reductions~\cite{Larsen17}.

\subsection{Locality-sensitive hashing}\label{sec:SimHash}

%Another research direction is investigating approximation solutions which trade accuracy for efficiency.
Locality-sensitive hashing (LSH)~\cite{Andoni08,Har12} is a key algorithmic primitive for similarity search in high dimensions due to the sublinear query time guarantee.
Several approaches exploit LSH to obtain sublinear solutions for approximate MIPS~\cite{Huang18,RangeLSH,Shrivastava14,SimpleLSH}.
We will review solutions based on SimHash~\cite{SimHash} since we can use SimHash for both similarity estimation and sublinear search for MIPS. 

\textbf{SimHash.}
Given a Gaussian random vector $\br \in \Rd$ whose elements are randomly drawn from the $N(0, 1)$, a SimHash function of $\bx$ is $h(\bx) = \sign\left( \dotxr \right)$. 
Denote by $ 0 \leq \thetaxq \leq \pi$ the angle between two vectors $\bx$ and $\bq$, the seminal work of Goemans and Williamson~\cite{Goemans} and Charikar~\cite{SimHash} show that
\begin{lemma}\label{lem:SimHash}
\begin{displaymath}
	\Pr{h(\bx) = h(\bq)} = 1 - \frac{\thetaxq}{\pi} = 1 - \frac{\arccos{\left(\dotxq/\|\bx\|\|\bq\|\right)}}{\pi} \, .
\end{displaymath}
\end{lemma}

Since $\arccos()$ is a monotonically decreasing function of $\dotxq$ when $\|\bx\| = \|\bq\| = 1$, SimHash is a LSH family for the inner product similarity on a unit sphere. 
%The SimHash-based algorithm is able to answer the $c$-approximate nearest neighbor search in $\BO {n^{1/c}}$ time.
Exploiting the binary representation of SimHash values, we can efficiently estimate the inner product $\dotxq$ with the fast Hamming distance computation using built-in functions of compilers.
We note that SimHash binary code can be seen as a quantization of Gaussian RP since it keeps the sign of each projected value.

\textbf{SimHash-based solutions for MIPS.}
Since the change of $\|\bq\|$ does not affect the result of MIPS, we can assume $\|\bq\| = 1$ without loss of generality.
Lemma~\ref{lem:SimHash} shows that the hash collision depends on both $\dotxq$ and $\|\bx\|$.
Therefore, by storing the 2-norm $\|\bx\|$, we can use SimHash for estimating $\dotxq$.
However, the dependency on such 2-norms makes MIPS more challenging to guarantee a sublinear query time.
Since inner product is not a metric, SimHash-based solutions have to convert MIPS to the nearest neighbor search by applying order preserving transformations to ensure that both data and query are on a unit sphere.

SimpleLSH~\cite{SimpleLSH} proposes asymmetric transformations : $\bq \mapsto \bq' = \{\bq, 0\}$ and $\bx \mapsto \bx' = \{\bx/M, \sqrt{1 - \|\bx\|^2 / M^2}\}$ where $M$ is the maximum 2-norm of all points in $\bX$.
Since the inner product order is preserved and $\|\bx'\| = \|\bq'\| = 1$, we can exploit SimHash for both similarity estimation and sublinear search for MIPS.

It is clear that the inner product values in the transformation space will be scaled down by a factor of $M$.
Furthermore, the top-$k$ inner product values are often very small compared to the vector norms in high dimensions.
This means that the MIPS values and the distance gaps between ``close'' and ``far apart'' points in the transformation space are arbitrary small.
Therefore, we need to use significantly large code length to achieve a reasonable search recall.
In addition, the standard $(l, L)$-parameterized (where $l$ is the number of concatenating hash functions and $L$ is number of hash tables) bucketing algorithm~\cite{Andoni08} demands a huge space usage (i.e. large $L$) to guarantee a sublinear query time.
In other words, the LSH performance will be degraded in the transformation space.

RangeLSH~\cite{RangeLSH} handles this problem by partitioning the data into several partitions and applying SimpleLSH on each partition.
While such distance gaps between ``close'' and ``far apart'' points in the LSH framework are slightly improved, the number of hash tables $L$ needs to be scaled proportional to the number of partitions to achieve a reasonable search recall.
Nevertheless, the subquadratic time and space cost of building LSH tables in order to guarantee a sublinear query time will be a bottleneck on many big data applications.

\begin{figure} [!t]
	\centering
	%\raggedleft
	\includegraphics[width=1\columnwidth]{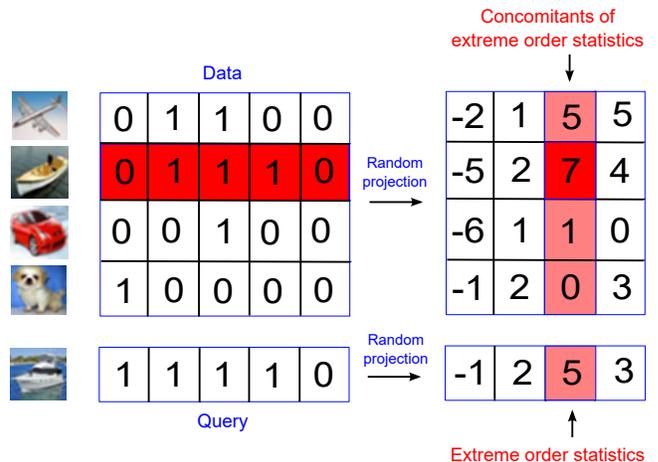}
	\caption{CEOs only uses the concomitants of extreme order statistics, i.e. the 3rd projected dimension corresponding to the maximum value of the query signature.
		The top-1 MIPS is the ``boat'' corresponding to the maximum value on the 3rd projection.
		The query time is extremely fast due to the pre-sorting the concomitants before querying.}
	\label{fig:CEOs}
\end{figure}

\subsection{Our contribution}

%LSH and traditional dimensionality reduction techniques for MIPS exploit Gaussian random projections~\cite{Kaban15,SimHash} to construct signatures (i.e. reduced dimensions) for both data and query points.
%Such methods use additional linear space to store $n$ signatures, and the query time is $\Omega(n)$ since we have to compare the query signature with $n$ data signatures.

The paper introduces a specific dimensionality reduction, called \emph{CEOs}, based the theory of concomitants of extreme order statistics.
While we also construct signatures for both data and query points using $D$ Gaussian random projections, we only use a small \emph{subset of $D$ dimensions} to estimate inner products.
This subset of size $s \ll D$ corresponds to random projections associated with the $s$ extreme values (i.e. maximum or minimum) of the query signature.
%random vectors closest and furthest to the query vector.
%This subset of size $s \ll d$ corresponds to the concomitants of extreme $s$th order statistics associated with the query signature.
%By exploiting the asymptotic properties of concomitants of extreme order statistics, we are able to show that such a small specific subset of data signatures contains sufficient information to estimate inner products and hence to solve top-$k$ MIPS.

The geometric intuition is that among $D$ Gaussian random projection vectors $\br_i, 1 \leq i \leq D$, the random projection associated with $\br_{(1)}$, the closest one to the query $\bq$, will preserve the best inner product order.
Since $\br_{(1)}$ is the closest vector to $\bq$, the value $\bq^\top \br_{(1)}$ is maximum among $D$ signature values $\bq^\top \br_i$.
Figure~\ref{fig:CEOs} shows a high level illustration of how CEOs works.
We note that ones can also use the random projection associated with $\br_{(D)}$, the furthest one to $\bq$ due to the symmetry of Gaussian distribution.

The technical difficulty is to provide a good estimator for $\dotxq$ given a small subset of projected values associated with the $s$ random projection vectors closest and furthest to $\bq$.
We observe that this subset of size $s \ll D$ is the concomitants of extreme $s$th normal order statistics associated with the query signature.
%By exploiting the asymptotic properties of these concomitants, we show that this small specific subset of data signatures contains sufficient information to estimate inner products and hence to solve top-$k$ MIPS.
Leveraging the theory of concomitants of extreme order statistics, we provide a surprisingly simple and asymptotically unbiased estimate for $\dotxq$.
That yields very fast and accurate MIPS solvers.

The key feature of CEOs is that we only need to use significantly small $s$ projected dimensions (often $s = 10$ in our benchmark) among $D$ signature dimensions.
Most importantly, in contrast to traditional dimensionality reductions, CEOs only uses the sign of a small subset of query signature (i.e. projections associated with maximum and minimum values) to estimate inner products.
This means that we can precompute and rank inner product estimators of all data \emph{before} querying to significantly reduce the query time.
We show that this cost of constructing indexes matches the lower bound of the $(1 + \epsilon)$-approximate MIPS in unit sphere~\cite{Andoni06}.

Under a mild condition, our theoretical analysis shows that we can achieve a sublinear query time for MIPS with search recall guarantees given an $\BO{(D/s)^s}$ indexing space  complexity. 
% under a weak assumption of data distribution and
To deal with such exponential time and space complexity, we introduce several CEOs variants which require $\TilO{dn}$ time and space for indexing and can answer top-$k$ MIPS with very high search recalls.
A summary of our contributions is as follows:
\begin{enumerate}
	\item We propose a novel dimensionality reduction method based on the theory of concomitants of extreme order statistics, named \emph{CEOs} for top-$k$ MIPS.
	We theoretically and empirically show that CEOs provides better estimate accuracy and hence higher top-$k$ MIPS recall than SimHash~\cite{SimHash} and  Gaussian random projections~\cite{Kaban15}.
	\item Inheriting from the asymptotic properties of concomitants of extreme $s$th order statistics, CEOs yields a \emph{sublinear} query time given $\BO{(D/s)^s}$ time and space complexity for building the index.
	The search recall can be theoretically guaranteed under a mild condition, and the exponential space usage is \emph{optimal} for MIPS on a unit sphere.
	%where $D$ is the signature dimensionality.
%	\item In spite of the exponential complexity, on the inferior choice that uses the concomitants of the smallest and largest order statistics, CEOs requires  less space and time complexity for building the index and returns significantly higher search recall than LSH-based solutions.
	%
	\item We propose practical CEOs variants, including \emph{sCEOs-TA} and \emph{coCEOs}, that build the index in $\TilO{dn}$ time.
	sCEOs-TA exploits the threshold algorithm~\cite{TA} for solving MIPS. 
	coCEOs is a data and dimension co-reduction technique and often outperforms sCEOs-TA on high search recall requirements.
	Especially, coCEOs can be viewed as a budgeted MIPS solver with a parameter $B = o(n)$ that answers top-$k$ MIPS in $o(n)$ time.
	\item Our proposed algorithms are very simple to implement and require few lines of codes.
%	Our empirical results confirm the efficiency of our proposed algorithms on many large-scale data sets.
%	 the sublinear CEOs achieves higher search recall than LSH bucket algorithms while the index construction is significantly faster and uses much less space.
	Empirically, on the inferior choice that uses the concomitants of the maximum order statistics, CEOs requires sublinear index space and outperforms LSH bucket algorithms regarding both efficiency and accuracy.
	Both sCEOs-TA and coCEOs outperform competitive MIPS solvers on answering top-10 MIPS and achieve at least 100x speedup compared to the bruteforce search with the accuracy at least 90\% on many real-world large-scale data sets.
\end{enumerate}

\section{Preliminaries} \label{sec:background}

%For any $d \times n$ matrix $\bX$, we let $\bxi$ be the  $i$-th row, $\bxj$ be the $j$-th column.
%We present the point set $\mX$ as a matrix ${\bX} \subset \Rnd$ where each point $\bx_i$ corresponds to the $i$-th row, i.e. $\bxi = (x_{i1}, \ldots, x_{id}) \in \Rd$, and the query point $\bq$ as a column vector $\bq = (q_{1}, \ldots, q_{d})^\top \subset \Rd$.

%For simplicity we assume that the matrix has full rank, the singular value decomposition (SVD) of $\bX \subset \Rnd$ is defined as 
%$$\bX = \bU \bS \bV^\top = \sum_{j = 1}^{d} {\sigma_j \buj \bvj^\top} \enspace ,$$
%where $\sigma_1, \ldots, \sigma_d$ are non-negative singular values sorted in descending order, $\buj$ and $\bvj$ are column vectors of $\bU$ and $\bV$, respectively.
%
%Let $\bsigma = \{\sigma_1, \ldots, \sigma_d\}$ be the row vector formed by the diagonal of $\bS$.
%We also denote $\bH$ and $\bD$ by the Hadamard matrix and the diagonal matrix with random values in $\{+1, -1\}$, respectively.
%We briefly review singular value decomposition (SVD), random projection, and random rotation methods. 

%For notation, we use lower-case fonts for scalars, upper-case fonts for random variables, bold lower-case fonts for vectors, and bold upper-case fonts for matrices.
%Let $[n] = \{1, 2, \ldots, n \}$, we use $i \in [n]$ and $j \in [d]$ to index the $i$-th row and the $j$-th column of ${\bX}$, respectively.
%We use $\|\bx\|$ as the 2-norm of $\bx$.

This section revises the background of bivariate normal distribution and concomitants of normal order statistics.

\subsection{Bivariate distribution}

Let $(U, V)$ be bivariate normal $N\left(\mu_U, \mu_V, \sigma_U^2, \sigma_V^2, \rho \right)$ where $\rho$ is the correlation coefficient, and $\mu_U, \mu_V, \sigma_U^2, \sigma_V^2 $ are the means and variances of $U$ and $V$, respectively. 
Let $(U_i, V_i)$ be a bivariate normal sample, we can write
\begin{align*}
V_i = \mu_V + \rho \frac{\sigma_V}{\sigma_U} (U_i - \mu_U) + \epsilon_i \, ,
\end{align*} 
where  $U_i$ and $\epsilon_i$ are mutually independent.
Furthermore, $\epsilon_i$ is normal with $\E{\epsilon_i} = 0, \Var{\epsilon_i} = \sigma_V^2 (1 - \rho^2)$.
%\begin{equation}\label{eq:bivariate}
%\begin{pmatrix}
%X \\
%Y
%\end{pmatrix} \sim N\left(
%\begin{pmatrix}
%0 \\
%0
%\end{pmatrix},
%\begin{pmatrix}
%\sigma_1^2 & \rho \\
%\dotxq & \sigma_2^2
%\end{pmatrix} \right).
%\end{equation}
%Denote $\rho = \dotxq / \|\bx\|\|\bq\|$ by the correlation coefficient of $Q$ and $X$.

~\cite[Section~4.7]{Bivariate_Book} shows that the conditional distribution of $V$ given $U = x$ is normal with the mean and variance as follows.

\begin{lemma}\label{lm:bivariate}
\begin{align*}
	\E{V | U = x} &= \mu_V + \rho \frac{\sigma_V}{\sigma_U}\left( x - \mu_U \right) \, , \\	
	\Var{V | U = x} &=  \sigma_V^2\left( 1 - \rho^2 \right) \, .
\end{align*}
\end{lemma}	

\subsection{Concomitants of normal order statistics}

Let $(U_1, V_1), (U_2, V_2), \ldots, (U_D, V_D)$ be $D$ random samples from a bivariate normal distribution $N\left(\mu_U, \mu_V, \sigma_U^2, \sigma_V^2, \rho \right)$.
We order these samples based on the $U$-value.
Given the $r$th order statistic $\osUr$, the $V$-value associated with $\osUr$ is called concomitant of the $r$th order statistic and denoted by $\cosVr$.
Let $\cosEr$ denote the specific $\epsilon_i$ associated with $\osUr$, we have the following equation:
\begin{align*}
\cosVr &= \mu_V+ \rho \frac{\sigma_V}{\sigma_U} (\osUr - \mu_U) + \epsilon_{[r]} \, .
\end{align*} 

The seminal work of David and Galambos~\cite{CEO_Paper} establishes the following properties of concomitants of normal order statistics.
\begin{lemma}\label{lm:COS}
\begin{align*}
\E{\cosVr} &= \mu_V + \rho \sigma_V \E{\frac{\osUr - \mu_U}{\sigma_U}} \, , \\
\Var{\cosVr} &= \sigma_V^2 \left( 1 - \rho^2 \right) + \sigma_V^2 \rho^2 \, \Var{\frac{\osUr - \mu_U}{\sigma_U}} \, .
\end{align*}	
\end{lemma}
	
In this work, we are interested in the concomitants of a few $r$th and $(D-r)$th order statistics where $r$ is small, for example $\cosVone$ and $\cosVd$.
The asymptotic distribution of these random variables with a sufficiently large $D$ has been studied in statistics during the last decades~\cite{CEO_Book,CEO_Paper}  with a considerable use in survival analysis.
%The following lemma shows the properties of $\cosYone$ and $\cosYd$.
	
\section{Theory of Concomitants of Extreme Order Statistics}\label{sec:theory}

Given two vectors $\bq, \bx \in \Rd$ and a random Gaussian vector $\br \in \Rd$ whose coordinates are randomly sampled from the normal distribution $N(0, 1)$, we let $Q = \bq^\top \br$ and $X = \bx^\top \br$.
For simplicity, we assume that $\|\bq\| = 1$ since it would not affect the efficiency and accuracy of our proposed MIPS solvers~\footnote{In practice, we do not need to normalize $\bq$ since all analysis will be scaled by a constant $\|\bq\|$.}.

It is well known that $Q \sim N(0, 1)$, $X \sim N(0, \|\bx\|^2)$.
More importantly, $Q$ and $X$ are normal bivariates from $N\left(0, 0, 1, \|\bx\|^2, \rho \right)$ where $\rho = \dotxq / \|\bx\|$~\cite{Li06,RP_Book}.

Let $(Q_1, X_1), (Q_2, X_2), \ldots, (Q_D, X_D)$ be $D$ random samples from $N\left(0, 0, 1, \|\bx\|^2, \rho \right)$.
We form the concomitants of normal order statistics by descendingly sorting these pairs based on $Q$-value.
The theory of concomitants of extreme order statistics studies the asymptotic behavior of the concomitants when $D$ goes to infinity.
In the following subsections, we will discuss how to leverage these asymptotic results to estimate inner products for MIPS.

\subsection{Concomitant of the extreme $\osQone$}\label{sec:first}

Let $\cosXone$ be the concomitant of the first (maximum) order statistic $\osQone$.
Applying Lemma~\ref{lm:COS}, we have the following properties of  $\cosXone$:
\begin{lemma}\label{lm:COS1}
	\begin{align*}
	\E{\cosXone} &= \rho \|\bx\| \E{\osQone} = \dotxq \,\, \E{\osQone}  \, , \\
	\Var{\cosXone} &= \|\bx\|^2 \left( 1 - \rho^2 \right) + \|\bx\|^2 \rho^2 \, \Var{\osQone} \\
	&= \|\bx\|^2 + \left( \dotxq \right)^2 \, \left(\Var{\osQone} - 1 \right) \, .
	\end{align*}	
\end{lemma}

We note that $\osQone$ is the largest variable among $D$ independent standard normal variables.
When $D$ is sufficiently large,~\cite[Chapter~8]{maxGauss_Book} and~\cite{maxGauss} show that $\E{\osQone} \rightarrow \sqrt{2 \log{D}}$ and $\Var{\osQone} \rightarrow 0$, and the rate of convergence is $\BO{1/\log{D}}$.
Using $\approxD$ as a proxy for asymptotic results with a sufficiently large $D$, we have
\begin{lemma}\label{lm:maxGauss}
	\begin{align*}
	\E{\cosXone} &\approxD \dotxq \sqrt{2 \log{D}} \,\, , \\
	\Var{\cosXone} &\approxD \|\bx\|^2 - \left( \dotxq \right)^2 \, .
	\end{align*}	
\end{lemma}

In order to use the concomitant of the first order statistics for estimating inner product value, we define $\Zcos = \cosXone / \sqrt{2 \log{D}}$.
It is straightforward that $\E{\Zcos} \approxD \dotxq$ and 
\begin{align}\label{eq:VarCOS}
\Var{\Zcos} \approxD \left( \|\bx\|^2 - \left( \dotxq \right)^2 \right)/2 \log{D} \, .
\end{align}

We note that the behavior of $\cosXone$ is consistent with the result of Lemma~\ref{lm:bivariate} where $x$ are the extreme values of standard normal distribution, e.g. around $\sqrt{2\log D}$.
While the variance is stable, the expectation is scaled up by a factor of $\sqrt{2\log D}$, which yields highly accurate estimates for MIPS.
For notational simplicity, we name the method using $\cosXone$ for estimating inner products as \emph{CEOs}. 

%Since the error induced by the asymptotic results of $\osQone$ only affects the variance estimate.
\textbf{Comparison with LSH:} We now compare the inner product estimators provided by SimHash and CEOs.
Define $Z = 1$ if $h(\bx) = h(\bq)$; otherwise 0.
By the LSH property, we have $\E{Z} = 1 - \thetaxq/\pi$ and $\Var{Z} = \left( \thetaxq / \pi \right) \left( 1 - \thetaxq/\pi \right) $.
Using the Taylor series with $\arccos{x} \approx \pi/2 - x$ for $-1 \leq x \leq 1$, we have
\begin{align*}
\E{Z} &= 1 - \frac{\thetaxq}{\pi} = 1 - \frac{\arccos{\left(\dotxq/\|\bx\|\right)}}{\pi} \approx \frac{1}{2} + \frac{\dotxq}{\pi \|\bx\|} \, ,\\
\Var{Z} &\approx \left( \frac{1}{2} - \frac{\dotxq}{\pi \|\bx\|} \right) \left( \frac{1}{2} + \frac{\dotxq}{\pi \|\bx\|} \right) = \frac{1}{4} - \frac{\left(\dotxq\right)^2}{\pi^2 \|\bx\|^2} \, .
\end{align*}
Define $\Zlsh = (Z - 1/2)\pi \|\bx\|$, it is clear that $\E{\Zlsh} \approx \dotxq$ and
\begin{align}\label{eq:VarLSH}
\Var{\Zlsh} = \pi^2 \|\bx\|^2 \Var{Z} \approx \frac{\pi^2}{4}\|\bx\|^2 - \left( \dotxq \right)^2 \, .
\end{align}

Equations~\ref{eq:VarCOS} and~\ref{eq:VarLSH} show that SimHash needs to use approximately $\frac{\pi^2}{2} \log{D} \approx 5 \log{D}$ bits to achieve a similar accuracy as CEOs with a sufficiently large $D$.
As elaborated in Section~\ref{sec:SimHash}, SimpleLSH applies SimHash for MIPS by scaling down all inner products with a normalization factor $M$ where $M$ is the maximum 2-norm of the data.
This means that SimpleLSH needs to use approximately $5M^2 \log{D}$ bits to have a similar accuracy as CEOs.

%Since we only need $\log{D}$ bits to encode the index of the concomitant, the message length of concomitant encoding is 5 times less the length of LSH codes when we want to estimate the inner products in distributed environments.

\textbf{Comparison with Gaussian RP:} 
We now bound the estimate error provided by $\cosXone$ and compare with the concentration bound provided by the Gaussian RP.
We investigate the asymptotic behavior of concomitants of extreme order statistics.
The theory of extreme order statistics~\cite{CEO_Paper} states that the distribution of the extreme case $\cosXone$ converges in probability to the normal distribution $N\left(\dotxq \sqrt{2 \log{D}}\, , \|\bx\|^2 - \left( \bx^\top \bq \right)^2 \right)$ when $D \rightarrow \infty$.
Recall that $\Zcos = \cosXone / \sqrt{2 \log{D}}$ and hence $\Zcos \approxD N\left( \dotxq \, , \frac{\|\bx\|^2 - \left( \dotxq \right)^2 }{2\log{D}} \right) \,$.

We will use the following Chernoff bounds~\cite{Chernoff} to bound the relative estimate error induced by $\cosXone$.
\begin{lemma}\label{lm:Chernoff}
	Let $Z \sim N(\mu, \sigma)$, for all $t \geq 0$ we have
	\begin{align*}
	\Pr{Z \geq \mu + t} &\leq \exp\left(-t^2 / 2\sigma^2\right) \, , \\
	\Pr{Z \leq \mu - t} &\leq \exp\left(-t^2 / 2\sigma^2\right) \, .
	\end{align*}	
\end{lemma}

Applying the Chernoff bounds for the random variable $\Zcos$, given any $\epsilon \geq 0$ we have:
\begin{align*}
&\Pr{|\Zcos - \dotxq| \geq \epsilon \, \dotxq} 
\leq 2\exp\left( \frac{-\epsilon^2 \log{D} \, \left( \dotxq \right)^2}{\|\bx\|^2 - \left( \dotxq \right)^2} \right)  \\
&= 2\exp\left(-\epsilon^2 \log{D} \cos^2{(\thetaxq)} / \sin^2{(\thetaxq)} \right) \\
&= 2D^{-\epsilon^2 \cos^2{(\thetaxq)} / \sin^2{(\thetaxq)}} \, \, .
\end{align*}

%This concentration bound is analog of Lemma~\ref{lm:JL} where $l = \BO{\log D}$.
When $D$ is sufficiently large, the error probability will be  arbitrarily small, and hence the estimate will be very accurate.
Using the asymptotic distribution of $\cosXone$, our concentration bound has the similar form of Lemma~\ref{lm:JL} with $l = 8\log{D} / \sin^2{(\thetaxq)}$ projections.
While in the worst case where $\thetaxq \rightarrow \pi/2$, Gaussian RP shares the similar bound as CEOs with $\BO{\log{D}}$.
However for the case of $\thetaxq \rightarrow 0$, CEOs offers a much tighter bound than Gaussian RP due to $\sin^2{(\thetaxq)} \rightarrow 0$.

\textbf{MIPS algorithms:} 
In general, compared to CEOs, SimHash and Gaussian RP can achieve similar performance for MIPS by using $\BO{\log{D}}$ bits and $\BO{\log{D}}$ random projections for constructing signatures, respectively.
However, after constructing signatures, these approaches have to perform $\BO n$ inner product estimation, which is clearly a computational bottleneck for MIPS.
In contrast, CEOs does not need this costly estimation step since we can sort all the points based on their projection values in advance.
At the query phase, after executing RP to compute the dimension index of $\osQone$, we can simply return $k$ point indexes corresponding to the top-$k$ concomitants associated with $\osQone$ as an approximate top-$k$ MIPS.
This key property makes CEOs more efficient than both Gaussian RP and LSH-based solutions on answering approximate MIPS.

Theoretically, the theory of concomitants of extreme order statistics requires $D \rightarrow \infty$.
However, we observe in practice that $D = 2^{10}$ suffices for many real-world data sets.
We present a brief experiment to compare the MIPS accuracy using $\cosXone$ called CEOs  with $D = 2^{10}$, SimHash, SimpleLSH and Gaussian RP.
We use the precision measure $P@k$ which computes the search recall of the top-$k$ answers provided by the estimation algorithms.
Figure~\ref{fig:CEO_Q1} shows the average measure $P@10$ and $P@20$ for $k = 10$ and $k = 20$ of these 4 methods on Nuswide over~100 queries.
The result is consistent with our analysis.
CEOs with $D = 2^{10}$ outperforms Gaussian RP with $l = 10$, and achieves higher accuracy than SimpleLSH with $l \leq 128$ and SimHash with $l < 32$ bits code.

\subsection{Concomitants of the extreme $\osQr$}\label{sec:second}

We observe that the Gaussian distribution is symmetric.
This means that we can use both $\cosXone$ and $\cosXd$ corresponding to the maximum $\osQone$ and minimum $\osQd$ order statistics for estimating $\dotxq$.
One natural question is: ``Can we use more concomitants and are they independent so that we can boost the accuracy with the standard Chernoff bounds?''.

Indeed, we have a positive answer by investigating the asymptotic independence of concomitants of extreme order statistics.
Let $s_0$ be a fixed positive integer and $D \rightarrow \infty$, the seminal work of David and Galambos~\cite{CEO_Paper} shows that the concomitants $\cosXr$ where $1 \leq r \leq s_0$ are asymptotically independent and  normal as follows:
\begin{align*}
\cosXr \approxD N\left(\bx^\top \bq \sqrt{2 \log{D}}, \|\bx\|^2 - \left( \bx^\top \bq \right)^2 \right) \, .
\end{align*}

By the symmetry of Gaussian distribution, we can use the concomitants $\cosXr$ and $X_{[D - r + 1]}$ where $1 \leq r \leq s_0$ to boost the accuracy of the estimator of $\dotxq$.
In other words, ones can view these $s = 2s_0$ concomitants as a specific dimension reduction since the sum of these values provides unbiased estimates of inner products.
%, but also to rank the $n$ inner product values.%The later property is essential to decrease the additional space usage and speed up MIPS, which will be presented in the next section.

By using the average of these $s$ concomitants to estimate $\dotxq$, we achieve an asymptotic unbiased estimator with the variance decreased by a factor of $1/s$.
Hence, the concentration bound is tighter by a factor of $1/\exp(s)$.

\begin{figure} [!t]
	\centering
	%\raggedleft
	\includegraphics[width=1\columnwidth]{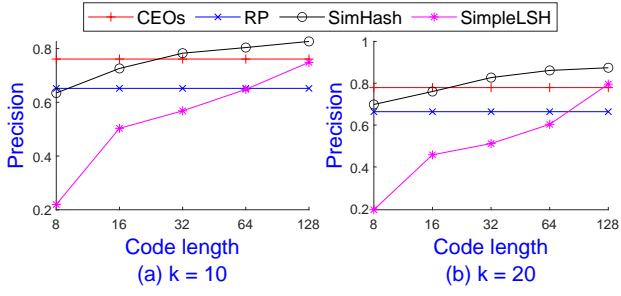}
	\caption{Precision of (a) top-10 MIPS and (b) top-20 MIPS on Nuswide using CEOs with $D = 2^{10}$, Gaussian RP 10 projections, and LSH variants with various the code lengths.}
	\label{fig:CEO_Q1}
\end{figure}

Figure~\ref{fig:CEO_Qs} (a) confirms our results on Nuswide by showing the average precision $P@10$ of CEOs on top-10 MIPS when varying $s$ and $D$.
%The CEOs uses the sum of $s$ concomitants with various settings of $s$ and $D$ on NUS-WIDE.
When fixing $s$ and $D$ is sufficiently large, e.g. $ \geq 512$, the precision marginally increases.
However, increasing $s$ yields a substantial rise of precision.
Especially, when $D = 512$ and $s = 4$, $P@10$ is nearly 90\%, which is higher than that of SimHash with 128 bits code shown in Figure~\ref{fig:CEO_Q1} (a). 
%Especially, $P@100$ reaches 75\% with just 10 dimensions whereas Gaussian RP returns less than 50\% with the same setting in Figure~\ref{fig:CEO_Q1}.

%\begin{figure} [!t]
%	\centering
%	%\raggedleft
%	\includegraphics[width=1\columnwidth]{Fig/Pa100_Mnist_Netflix.eps}
%	\caption{Accuracy $P@100$ of top-10 MIPS when (a) fixing $s = 4$ and varying $D$, and (b) fixing $D = 512$ and varying $s$ on Mnist and Netflix.}
%	\label{fig:CEO_Qs}
%\end{figure}

%\begin{figure*} [!t]
%	\centering
%	%\raggedleft
%	\includegraphics[width=1\textwidth]{Fig/Pat100_Precision_Recall_Netflix_Mnist.eps}
%	\caption{Accuracy $P@100$, Precision-Recall curve of top-10 MIPS on Netflix and Mnist when varying $l$.}
%	\label{fig:CEO_LSH_Est}
%\end{figure*}

\section{Efficient Algorithms for top-$k$ MIPS}\label{sec:algorithm}

As illustrated in Section~\ref{sec:second}, we can view the concomitants of extreme order statistics as a specific dimension reduction.
The number of concomitants associated with the extreme $s$th order statistics is the number of reduced dimensions.
Though $s$ is often very small in practice, the linear time of estimating $n$ inner products will be the computational bottleneck of MIPS.
This section will describe how to exploit key properties of concomitants of extreme $s$th order statistics to achieve $o(n)$ MIPS solvers.

We recall the notation that we have $\bq, \bx, \by \in \Rd$, $\|\bq\| = 1$, and a random Gaussian vector $\br \in \Rd$ whose coordinates are sampled from $N(0, 1)$.
We let $Q = \bq^\top \br$, $X = \bx^\top \br$, and $Y = \by^\top \br$.
%We also have $Q \sim N(0, 1)$, $X \sim N(0, \|\bx\|^2)$, and $Y \sim N(0, \|\by\|^2)$. 
We have $Q$ and $X$ are normal bivariates from $N\left(0, 0, 1, \|\bx\|^2, \rhox \right)$ where $\rhox = \dotxq / \|\bx\|$. 
$Q$ and $Y$ are normal bivariates from $N\left(0, 0, 1, \|\by\|^2, \rhoy \right)$ where $\rhoy = \dotyq / \|\by\|$. 
 
Let $\cosXr$ and $\cosYr$ be the concomitants of the $\osQr$ in $D$ random samples from the bivariate normal distribution $N\left(0, 0, 1, \|\bx\|^2, \rho_{\bx} \right)$ and $N\left(0, 0, 1, \|\by\|^2, \rho_{\by} \right)$, respectively.
We note that $\cosXr$ and $\cosYr$ are not independent due to the link to $\osQr$.

For simplicity, we only consider $\cosXone$ and $\cosYone$ and denote $\rhoone$ be the correlation between $\cosXone$ and $\cosYone$.
Let $\tau_1 = \dotxq$, $\tau_2 = \dotyq$ and assume $\tau_1 > \tau_2$.
By the asymptotic property of concomitants of extreme order statistics, we have
\begin{align*}
\cosXone &\approxD N\left(\tau_1 \sqrt{2 \log{D}}, \|\bx\|^2 - \tau_1^2 \right) \, , \\
\cosYone &\approxD N\left(\tau_2 \sqrt{2 \log{D}}, \|\by\|^2 - \tau_2^2 \right) \, .
\end{align*}
Let $\sigmaone^2 = \|\bx\|^2 + \|\by\|^2 - \tau_1^2 - \tau_2^2 + 2 \rhoone \sqrt{\|\bx\|^2 - \tau_1^2} \sqrt{\|\by\|^2 - \tau_2^2}$.
Since the sum of two Gaussian variables is Gaussian, we have
\begin{align*}
\cosYone - \cosXone \approxD N\left( \left(\tau_2 - \tau_1 \right) \sqrt{2 \log{D}}, \sigmaone^2\right) \, .
\end{align*}
Applying Lemma~\ref{lm:Chernoff} for the Gaussian variable $\cosYone - \cosXone$, we have
\begin{align*}
&\Pr{\cosYone - \cosXone \geq 0} \\
&= \Pr{\cosYone - \cosXone \geq \E{\cosYone - \cosXone} + \left(\tau_1 - \tau_2\right) \sqrt{2 \log{D}}} \,  \\
&\leq e^{-\left(\tau_1 - \tau_2\right)^2 \log{D} /  \sigmaone^2} = D^{-\left(\tau_1 - \tau_2\right)^2 / \sigmaone^2} \,.
\end{align*}
Since $-1 \leq \rhoone \leq 1$, we have $\sigmaone^2 \leq \left( \sqrt{\|\bx\|^2 - \tau_1^2} + \sqrt{\|\by\|^2  - \tau_2^2} \right)^2$.
Therefore,
\begin{align}\label{eq:CEO}
\Pr{\cosYone - \cosXone \geq 0} &\leq D^{-\left(\tau_1 - \tau_2\right)^2 / \left( \sqrt{\|\bx\|^2 - \tau_1^2} + \sqrt{\|\by\|^2  - \tau_2^2} \right)^2} \, .
\end{align}

\begin{figure} [!t]
	\centering
	%\raggedleft
	\includegraphics[width=1\columnwidth]{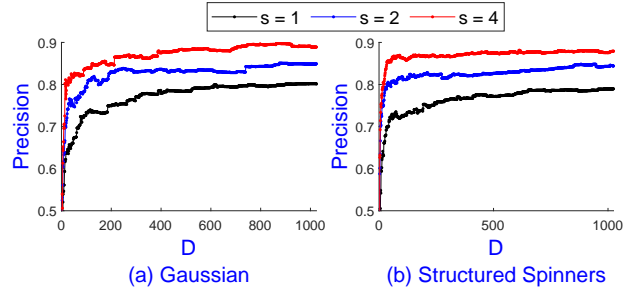}
	\caption{Top-10 MIPS precision on Nuswide with $s = \{1, 2, 4\}$ and varying $D$ using: (a) Gaussian RP and (b) Structured Spinners to simulate Gaussian RP.}
	\label{fig:CEO_Qs}
\end{figure}

Let $\alpha_{\by} = \left(\tau_1 - \tau_2\right) / \left( \sqrt{\|\bx\|^2 - \tau_1^2} + \sqrt{\|\by\|^2  - \tau_2^2} \right) > 0$.
When $\alpha^2_{\by}$ is not very small and $D$ is sufficiently large, $\Pr{\cosYone \geq \cosXone} \leq \delta$ for any $0 < \delta < 1$.
%the probability that that $\cosYone$ is ranked lower than $\cosXone$ will be small.
%we can get $\Pr{\cosXone - \cosYone \geq 0} \geq 1 - \delta$ for any $0 < \delta < 1$.
In order words, $\cosXone$ is ranked higher than $\cosYone$  with high probability in the sorted list associated with $\osQone$.
This observation is the key property which makes CEOs more efficient than competitive MIPS solvers.

Due to the asymptotic independence and normal distribution of $s = 2s_0$ concomitants associated with $\osQr$ and $Q_{(D - r + 1)}$ where $1 \leq r \leq s_0$, we can use the sum of these $s$ concomitants as an inner product estimate.
Note that for $Q_{(D - r + 1)}$, we have to reverse the sign due to the symmetry of normal distribution.
Since the variance of the sum of $2s_0$ independent random variables is decreased by a factor of $1/2s_0$, we have
\begin{align}\label{eq:sCEO}
\Pr{\sum_{r = 1}^{s_0}\left(Y_{[r]} - Y_{[D - r + 1]} \right) \geq \sum_{r = 1}^{s_0}\left(X_{[r]} - X_{[D - r + 1]} \right)} \leq D^{-2s_0\alpha_{\by}^2} \, .
\end{align}

%By choosing $s_0 = \log{n}$, the probability above is bounded by $n^{-2\log{D}\alpha^2_{\by}}$.
Assume that $\alpha^2_{\by}$ is not very small, i.e. $\alpha_{\by}^2 \geq c / s_0$ where $c> 1$.
Given a sufficiently large $n$, $D = n^{1/c}$ suffices for asymptotic properties of concomitants associated with the $s$th order statistics.
With $D = n^{1/c}$, the probability in Equation~\ref{eq:sCEO} is bounded by $1/n^2$.
Applying the union bound, we state our main result as follows:

%We now bound the failure probability when using the sum of $s$ concomitants associated with $s$th extreme order statistics for answering MIPS.
%
%Denote $\alpha = \argmin_{\by \in \bX \setminus \bx} \alpha_{\by}$, then for all $\by \in \bX \setminus \bx$ we have
%%
%\begin{align}\label{eq:sCEO}
%\Pr{\sum_{r = 1}^{s_0}\left(Y_{[r]} + Y_{[D - r + 1]} \right) \geq \sum_{r = 1}^{s_0}\left(X_{[r]} + X_{[D - r + 1]} \right)} \leq D^{-2s_0\alpha^2} \, .
%\end{align}

\begin{theorem}\label{thm:main}
	Assume $\bx$ is the top-1 MIPS given the query $\bq$.
	Let $\alpha_{\by} = \left(\dotxq - \dotyq\right) / \left( \sqrt{\|\bx\|^2 - (\dotxq)^2} + \sqrt{\|\by\|^2  - (\dotyq)^2} \right)$ for any $\by \in \bX \setminus \bx$. 
	Given a sufficiently large $n$ and a constant $c > 1$, we assume that $\alpha_{\by}^2 > c / s_0$.
	By choosing $D = n^{1/c}$, for \emph{all} $\by \in \bX \setminus \bx$, we have: 
	\begin{align*}
	\Pr{\sum_{r = 1}^{s_0} \left(X_{[r]} - X_{[D - r + 1]} \right) \geq \sum_{r = 1}^{s_0}\left(Y_{[r]} - Y_{[D - r + 1]} \right)} \geq 1 - 1/n \, .
	\end{align*}
\end{theorem}

Our result does not hold on the data sets where inner products between all data points and the query are almost similar. 
This is due to the fact that $\alpha_{\by} \rightarrow 0$ and the assumption $\alpha^2_{\by} > c /s_0$ does not hold.
%In such worst case, there is no known algorithms for solving exactly MIPS in truly sub-linear time.
%In fact, if there exists such an algorithm, the Strong Exponential Time Hypothesis (SETH)~\cite{SETH}, a fundamental conjecture in computational complexity theory, is wrong~\cite{Chen19,Williams14}.
Nevertheless, real-world data sets rarely have such properties and our empirical results show that CEOs with $D = 1024$ works very efficiently, as shown in Figure~\ref{fig:CEO_Q1} and~\ref{fig:CEO_Qs}.

\subsection{Optimality for $(1 + \epsilon)$-MIPS on a unit sphere}

This subsections discusses the optimality of CEOs on answering the $(1 + \epsilon)$-approximate MIPS on a unit sphere given a small $\epsilon$.
Given a point set $\mX \subset \Rd$ of size $n$ and a query point $\bq \in \Rd$ on a unit sphere, it is clear that $\argmax_{\bx \in \mX}{\bx^\top \bq} = \argmin_{\bx \in \mX}{\| \bx - \bq \| }$.
Therefore, we investigate the result of theorem~\ref{thm:main} for a decision version of the approximate nearest neighbor search over the Euclidean space on a unit sphere.
The decision version is to build a data structure that answers $(1 + \epsilon)$-approximate near neighbor search as follows.
\begin{itemize}
	\item If there is $\bx \in \mX$ such that $\| \bx - \bq \| \leq r$, answer YES.
	\item If there is no $\bx \in \mX$ such that $\| \bx - \bq \| \leq (1 + \epsilon)r$, answer NO.
\end{itemize}

Andoni et al.~\cite{Andoni06} analyze the lower bound of the decision problem given a constant query time (measured by the number of probes to the data structure) on the Euclidean space.
In particular, any algorithm uses a constant number of probes to the data structure must use $n^{\Omega(1/\epsilon^2)}$ space.
Since a high search recall demands very small $\epsilon$, this result presents a fundamental difficulty for any practical data structures that can answer the nearest neighbor search very accurate and fast. 

An upper bound for $(1 + \epsilon)$-nearest neighbor search over the Hamming space is proposed in~\cite{Kushilevitz00} that use $n^{\Osymbol(1/\epsilon^2)}$.
This work complements that result by showing a data structure which achieves a constant query time and uses $n^{\Osymbol(1/\epsilon^2)}$ over the Euclidean space on a unit sphere.
We show it by deriving and bounding the value $\alpha_{\by}$ on Theorem~\ref{thm:main} where $\| \bx - \bq \| \leq r$ and $\| \by - \bq \| \geq (1 + \epsilon)r$ as follows.
\begin{align*}
\alpha_{\by} &= \frac{\dotxq - \dotyq}{\sqrt{1 - (\dotxq)^2} + \sqrt{1 - (\dotyq)^2} }\\
			&= \frac{\left(1 - r^2/2 \right)  - \left(1 - (1 + \epsilon)^2r^2/2 \right)}{\sqrt{1 - (1 - r^2/2)^2} + \sqrt{1 - \left(1 - (1 + \epsilon)^2r^2/2 \right)^2}} \\
			&= \frac{\left( 2\epsilon + \epsilon^2 \right) r^2/2}{\sqrt{r^2 - r^4/4} + \sqrt{(1 + \epsilon)^2r^2 - (1 + \epsilon)^4r^4/4}} \\			
			&= \frac{\epsilon \left(\epsilon + 2 \right) }{\sqrt{4/r^2 - 1} + (1 + \epsilon)\sqrt{4/ r^2 - (1 + \epsilon)^2}}
\end{align*}
Hence, 
\begin{align*}
\frac{\epsilon \left(\epsilon + 2 \right) }{(2 + \epsilon)\sqrt{4/r^2 - 1}} \leq \, & \alpha_{\by} \, \leq \frac{\epsilon \left(\epsilon + 2 \right) }{(2 + \epsilon)\sqrt{4/r^2 - (1 + \epsilon)^2}} \\\frac{\epsilon }{\sqrt{4/r^2 - 1}} \leq \, & \alpha_{\by} \, \leq \frac{\epsilon}{\sqrt{4/r^2 - (1 + \epsilon)^2}}
\end{align*}

On a unit sphere and for a fixed distance $0 \leq r \leq 2$, we have $\alpha_{\by} = \Theta(\epsilon)$.
From Equation~\ref{eq:sCEO}, we have to use $D = n^{1/\alpha_{\by}^2} = n^{\Theta(1/\epsilon^2)}$ to answer the decision version of $(1 + \epsilon)$-approximate near neighbor search with high probability.
This means that CEOs can precompute the results of $(1 + \epsilon)$-approximate near neighbor search for all queries using $n^{\Theta(1/\epsilon^2)}$ space and answer the query with a constant probes.
This matches the lower bound on the space usage provided by~\cite{Andoni06}.

\begin{algorithm}[!t]
	\caption{1CEOs}
	\label{alg:CEOs} 									
	\begin{algorithmic} [1]
		%-----------------------------------------
		\Function{Indexing} {data matrix $\bX_{d \times n}$, random Gaussian matrix $\bR_{D \times d}$}
		%-----------------------------------------
		\State Project $\bX$ into $D$ dimensions by computing $\bX' = \bR\bX$		
		\State For each dimension $j \in [D]$, partially sort $\bX'_j$ to add  top-$b$ indexes with the largest values to the list $L_j$
		\State \Return $D$ lists $L_j$ where $j \in [D]$ as our index
		\normalsize
		\EndFunction
		\algrule
		%-----------------------------------------
		\Procedure{Querying} {query point $\bq \in \Rd$, data matrix $\bX_{d \times n}$, random Gaussian matrix $\bR_{D \times d}$, our index}
		%-----------------------------------------
		\State Project $\bq$ into $D$ dimensions by computing $\bq' = \bR \bq$		
		\State Compute dimension $j_{(1)}$ of the maximum value of $\bq'$
		\State Post-processing: Compute $b$ inner products from the list $L_{j_{(1)}}$ and return top-$k$ points with the largest value
		%		\RETURN { $D$ lists $L_j$, $1 \leq j \leq D$, as our indexing}
		%
		\normalsize
		\EndProcedure
	\end{algorithmic}
\end{algorithm}

\subsection{MIPS using concomitants of $\osQone$}
Consider that we have a small budget of $b$ inner product computations for post-processing to achieve higher MIPS accuracy.
For each projected dimension, we only need to maintain top-$b$ largest concomitants among $n$ concomitants corresponding to $n$ points.
For querying, we only use these $b$ concomitants associated with $\osQone$.
Since any MIPS solver can use post-processing, we will consider $b = \BO 1$ to simplify the analysis.
Algorithm~\ref{alg:CEOs} shows how CEOs with concomitants of $\osQone$, named \emph{1CEOs}, works.

%\begin{algorithm}[!t]
%	\caption{CEOs: Indexing}
%	\label{alg:CEOs_ND} 									
%	\begin{algorithmic} [1]
%		%-----------------------------------------
%		\REQUIRE { Data matrix $\bX_{d \times n}$, random Gaussian matrix $\bR_{D \times d}$}
%		\ENSURE { Indexing structure }
%		%-----------------------------------------
%		\STATE { Random project $n$ points into $D$ dimensions by computing $\bR\bX$}		
%		\STATE { For each dimension $j$, sort $n$ values and keep top-$b$ point indexes with the largest values in the list $L_j$ }
%		%
%		\RETURN { $D$ lists $L_j$, $1 \leq j \leq D$, as our index}
%		%
%		\normalsize
%	\end{algorithmic}
%\end{algorithm}
%%
%\begin{algorithm}[!t]
%	\caption{CEOs: Querying}
%	\label{alg:CEOs_ND} 									
%	\begin{algorithmic} [1]
%		%-----------------------------------------
%		\REQUIRE { Query vector $\bq \in \Rd$, random Gaussian matrix $\bR_{D \times d}$, our index}
%		\ENSURE { Approximate MIPS }
%		%-----------------------------------------
%		\STATE { Random project $\bq$ points into $D$ dimensions by computing $\bR \bq$}		
%		\STATE { Compute dimension $j_1$ corresponding to the maximum $\osQone$ }
%		\STATE { Post-processing: Compute $b$ inner products from the list $L_{j_1}$ and return top-$k$ points with the largest value }
%%		\RETURN { $D$ lists $L_j$, $1 \leq j \leq D$, as our indexing}
%		%
%		\normalsize
%	\end{algorithmic}
%\end{algorithm}

\textbf{Complexity:} It is clear that building the index takes $\BO{dDn}$ time. 
%Especially, the memory use of our index is only $\BO{bdD}$, which is small enough for many data stream applications.
The index space is only $\BO{dD}$ since each of $D$ dimensions stores $b$ points, and the query time is $\BO{dD}$.
When $D = o(n)$, 1CEOs answers MIPS in both sublinear space and time.

\textbf{Error analysis:} Equation~\ref{eq:CEO} indicates that, given a sufficiently large $D$, $\cosXone$ corresponding to the top-$k$ MIPS tends to be ranked at the top positions on the list $L_{j_{(1)}}$ corresponding to $\osQone$.
We note that by simply increasing $b$ inner products in post-processing, we can achieve higher accuracy of top-$k$ MIPS since we allow larger gap $\cosYone - \cosXone$.
While we cannot theoretically guarantee the performance of 1CEOs, our empirical results show that 1CEOs outperforms the sublinear LSH bucket algorithm regarding both space and time of indexing and querying for MIPS.

%\textbf{Comparison with LSH-based schemes:} While we can use LSH for both estimation with binary code and sublinear search with multiple hash tables, these strategies result in either linear query time  or subquadratic time of constructing the index.
%In contrast, CEOs requires near-linear time for building the index, and sublinear time for both querying and updating the index.

%\textbf{Comparison with Gaussian RP:} After projecting to $l$ dimensions, the standard Gaussian RP needs to estimate $n$ inner products in $\BO{nl}$ time.
%In contrast, CEOs only requires $\BO{bd}$ computation from the $\osQone$ dimension.
%Therefore, while CEOs and Gaussian RP share similar concentration bounds, CEOs clearly outperforms Gaussian RP on answering MIPS.

%We note here that we can achieve even stronger bound if considering the larger gap, i.e. $\cosYone - \cosXone \geq \left(\tau_1 - \tau_2\right) \sqrt{2 \log{n}}$ or setting $D = \BO{n}$ to guarantee the error bound for \emph{all} $n$ pairs.
%Furthermore, due to the asymptotic independence and the symmetry of the normal order statistics, we can use the $2m$ concomitants $\cosXr$ and $X_{[D - r]}$ where $m$ is a fixed integer and $1 \leq r \leq m$ to amplify the probability of success in at least one trial to $1 - 1/2m$.

\subsection{Sublinear MIPS using concomitants of $\osQr$}

Since we can use $s = 2s_0$ concomitants associated with $\osQr$ and $Q_{(D + 1 - r)}$ where $1 \leq r \leq s_0$ to boost the estimate accuracy, Theorem~\ref{thm:main} shows that we can have a sublinear MIPS with guarantees.

Since the estimation does not use the values of query signatures (except the dimension order), we can precompute and sort \emph{all} possible estimates before querying. 
%aggregate and sort all possible combinations between these $s$ concomitants.
For example, in order to use $\cosXone$ and $\cosXd$ for estimating $\dotxq$, after executing $D$ random projections, we can precompute the difference of all pairs of dimensions and sort the data based on these difference values.
We compute the difference $\cosXone - \cosXd$ because of the symmetry of normal bivariate distribution.
Algorithm~\ref{alg:sCEOs}, named \emph{sCEOs}, generalizes this observation to exploit concomitants of extreme $s$th order statistics for MIPS.

\begin{algorithm}[!t]
	\caption{sCEOs}
	\label{alg:sCEOs} 									
	\begin{algorithmic} [1]
		%-----------------------------------------
		\Function{Indexing} {data matrix $\bX_{d \times n}$, random Gaussian matrix $\bR_{D \times d}$}
		%-----------------------------------------
		\State Project $\bX$ into $D$ dimensions by computing $\bX' = \bR\bX$	
		\State For each pair of sets $I$ and $J$, each containing distinct $s_0$ dimensions among $D$ dimensions and $I \cap J = \emptyset$, partially sort $\sum_{i \in I}\bX'_{i} - \sum_{j \in J}\bX'_{j}$ to add top-$b$ indexes with the largest values to the list $L_{IJ}$
		\State \Return $\BO{\left(D/s\right)^s}$ lists $L_{IJ}$ where $I$ and $J$ correspond to concomitants of the $s_0$ maximum and minimum order statistics
		\normalsize
		\EndFunction
		\algrule
		%-----------------------------------------
		\Procedure{Querying} {query point $\bq \in \Rd$, data matrix $\bX_{d \times n}$, random Gaussian matrix $\bR_{D \times d}$, our index}
		%-----------------------------------------
		\State Project $\bq$ into $D$ dimensions by computing $\bq' = \bR \bq$		
		\State Compute the sets $I$ and $J$ of $s_0$ dimensions corresponding to $s_0$ maximum and minimum values of $\bq'$
		\State Post-processing: Compute $b$ inner products from the list $L_{IJ}$ and return top-$k$ points with the largest value
		%		\RETURN { $D$ lists $L_j$, $1 \leq j \leq D$, as our indexing}
		%
		\normalsize
		\EndProcedure
	\end{algorithmic}
\end{algorithm}

\textbf{Complexity and error analysis: } Building the index takes $\BO{\left(D/s\right)^s n}$ time and $\BO{\left(D/s\right)^s + dn}$ space.
This is because we have $\binom{D}{s_0}$ different sets $I$ and each of $I$ has $\binom{D-s_0}{s_0}$ different sets $J$.
sCEOs has $\BO{dD}$ query time as similar as 1CEOs.
%In general, 2CEOs returns higher accuracy than CEOs with the same querying time but requires more indexing construction time and space.
Setting $D = n^{1/c}$ for any $c > 1$, Theorem~\ref{thm:main} states that sCEOs can have a \emph{sublinear} query time given that the assumption $\alpha^2_{\by} > c/s_0$ holds for all $\by$.
Furthermore, sCEOs returns exact top-$k$ MIPS with probability at least $1 - 1/n$.
%one can simply generalize this procedure for $s$ concomitants of extreme order statistics.
%This variant is called by \emph{sCEOs} and answers MIPS in sublinear time, i.e. $\BO{dD + db}$ where $D \ll n$.
Figure~\ref{fig:CEO_Qs} (a) demonstrates that with $s = 4$, $D = 512$ and $b = 10$, sCEOs can achieve nearly 90\% search recall for top-10 MIPS on Nuswide even without post-processing.

\textbf{Performance simulation: } One of the nice features of sCEOs is that we can compute the sCEOs search recall by implementing it as a specific dimensionality reduction.
We name this approach as \emph{sCEOs-Est} since it estimates $n$ inner products by the sum of concomitants of extreme $s$th order statistics of the query signature. 
While sCEOs-Est estimates $n$ inner products in $\BO{sn}$ time, it still runs significantly faster than both Gaussian RP and bruteforce search since addition operators are often much faster than multiply-add operators.
More important, we can use sCEOs-Est to simulate the sublinear search performance of sCEOs \emph{before} implementing it.
This property makes sCEOs very useful in practice while LSH bucket algorithms do not have.

\subsection{TA algorithm using concomitants of $\osQr$}

While the querying time is sublinear, the space and time complexity for building the index of sCEOs blow up by a factor of $(D/s)^s$, which will be the computational bottleneck of many big data applications.
Therefore, we might need to use sCEOs-Est for large data sets.

We observe that sCEOs-Est estimates inner product values by the sum of $s$ positive concomitants and extracts top-$b$ indexes with the largest estimates for post-processing.
%instead of precomputing all possible estimates in $\BO{D^s}$ time
Therefore, ones can speed up sCEOs-Est by using the well-known threshold algorithm (TA)~\cite{TA}.
Algorithm~\ref{alg:sCEOs-TA}, named \emph{sCEOs-TA}, shows how we can exploit the TA algorithm for speeding up sCEOs-Est.

\begin{algorithm}[!t]
	\caption{sCEOs-TA}
	\label{alg:sCEOs-TA} 									
	\begin{algorithmic} [1]
		%-----------------------------------------
		\Function{Indexing} {data matrix $\bX_{d \times n}$, random Gaussian matrix $\bR_{D \times d}$}
		%-----------------------------------------
		\State Project $\bX$ into $D$ dimensions by computing $\bX' = \bR\bX$		
		\State For each dimension $j \in [D]$, sort $\bX'_j$ and keep it together with the point indexes on the list $L_j$
		\State \Return $D$ lists $L_j$ where $j \in [D]$ as our index
		\normalsize
		\EndFunction
		\algrule
		%-----------------------------------------
		\Procedure{Querying} {query point $\bq \in \Rd$, data matrix $\bX_{d \times n}$, random Gaussian matrix $\bR_{D \times d}$, our index}
		%-----------------------------------------
		\State Project $\bq$ into $D$ dimensions by computing $\bq' = \bR \bq$		
		\State Compute $s = 2s_0$ dimensions corresponding to $s_0$ maximum and minimum values of $\bq'$
		\State Apply the TA algorithm over these $s$ dimensions to compute top-$b$ indexes with the largest inner product estimators
		\State Post-processing: Compute $b$ inner products provided by the TA algorithm and return top-$k$ points with the largest value
		%		\RETURN { $D$ lists $L_j$, $1 \leq j \leq D$, as our indexing}
		%
		\normalsize
		\EndProcedure
	\end{algorithmic}
\end{algorithm}

\textbf{Complexity: } sCEOs-TA builds the index in $\BO{dDn + Dn\log{n}}$ time, which is slightly larger than $\BO{dDn}$ time of sCEOs-Est, but uses the same $\BO{Dn + dn}$ space.
Unfortunately, we could not state the query time complexity of sCEOs-TA.
In practice, sCEOs-TA provides the same accuracy as sCEOs-Est due to the same computation of top-$b$ estimators for post-processing.
However, our empirical results show that sCEOs-TA often runs 5 -- 10 times faster than sCEOs-Est, especially when $s$ and $b$ are small.
It is worth noting that sCEOs-TA can be used to speed up the index construction of sCEOs since it computes exactly top-$b$ inner product estimates for each set of $s$ dimensions.

\subsection{Co-reduction using concomitants of $\osQr$} 

%Since we can use a few dimensions corresponding to extreme order statistics $\osQr$, sCEOs-Est can be seen as a dimensionality reduction.
While sCEOs-TA can speed up sCEOs-Est given a similar time and space complexity for constructing the index, we could not bound its running time.
From Equation~\ref{eq:CEO}, we observe that we do not need to keep the whole data in the sorted list $L_{(r)}$ associated with the dimension of $\osQr$ for answering MIPS.
By exploiting this property, we propose \emph{coCEOs}, a co-reduction method that keeps a fraction of data in our index and uses a small number of dimensions for answering MIPS.

Since one dimension can associate with the maximum or minimum order statistics, we need to keep both top-$m$ point indexes with the largest and smallest values.
Therefore, coCEOs can be seen as a compressed data structure that keeps the top-$2m$ points on each dimension $j \in [D]$.
When the query comes, we will choose the smallest or largest top-$m$ indexes depending on the rank of $\osQr$.
We observe that the larger inner product the point has, the more frequent it should be listed on these top-$2m$ points, and hence the larger estimate it has.
%Therefore, the frequency of the points on the compressed data structure is a good estimate to rank inner products.
We propose coCEOs to compute partial estimate of the inner products in sCEOs-based data structure, as shown in Algorithm~\ref{alg:coCEOs} . 

\begin{algorithm}[!t]
	\caption{coCEOs}
	\label{alg:coCEOs} 									
	\begin{algorithmic} [1]
		%-----------------------------------------
		\Function{Indexing} {data matrix $\bX_{d \times n}$, random Gaussian matrix $\bR_{D \times d}$}
		%-----------------------------------------
		\State Project $\bX$ into $D$ dimensions by computing $\bX' = \bR\bX$		
		\State For each dimension $j \in [D]$, partially sort $\bX'_j$ to add top-$m$ indexes with the largest and smallest values to the lists $L_j$ and $S_j$, respectively
		\State \Return $2D$ lists $L_j$ and $S_j$ where $j \in [D]$ as our index
		\normalsize
		\EndFunction
		\algrule
		\Procedure{Querying} {query point $\bq \in \Rd$, data matrix $\bX_{d \times n}$, random Gaussian matrix $\bR_{D \times d}$, our index}
		%-----------------------------------------
		\State Project $\bq$ into $D$ dimensions by computing $\bq' = \bR \bq$	
		\State Compute the sets $I$ and $J$ of $s_0$ dimensions corresponding to $s_0$ maximum and minimum values of $\bq'$	
%		\State Compute $s = 2s_0$ dimensions corresponding to the $s_0$ largest and smallest values of $\bq'$
		\State For each $i \in I$ and $j \in J$, scan all points in lists $L_i$ and $S_{j}$, and update their partial estimates in the histogram
		\State Retrieve top-$b$ points with the largest partial estimates
		\State Post-processing: Compute $b$ inner products from these points and return top-$k$ points with the largest value
		%		\RETURN { $D$ lists $L_j$, $1 \leq j \leq D$, as our indexing}
		%
		\normalsize
		\EndProcedure
	\end{algorithmic}
\end{algorithm}

\textbf{Complexity: } While the space usage of our index is $\BO{mD + dn}$, the construction time is $\BO{dDn + Dn\log{m}}$.
By using a hash table to maintain the histogram of partial estimates, the query time of coCEOs is $\BO{dD + ms}$. 
Notable, coCEOs requires linear space but still runs in sublinear time when $ms = o(n)$.
%Notably, coCEOs has sublinear query time when $m = o(n)$.
%Since we can govern the query time of coCEOs, we can use it for the budgeted MIPS~\cite{dWedge} in recommendation system.

\textbf{Practical setting:} While sCEOs-Est sums the $s$ dimensions associated with $\osQr$ for $n$ points, coCEOs computes partial estimates for the points that are likely to be the top-$k$ MIPS.
Hence, coCEOs can exploit a larger number of extreme order statistics (i.e. larger number of dimensions) compared to sCEOs-Est.
%In other words, $s_{coCEOs} > s_{sCEOs}$.
Let $s' > s$ be the number of dimensions used by coCEOs.
In order to govern the running time of coCEOs, we set a budget of $B$ samples and select $B/s'$ points on each of $s'$ dimensions associated with $\osQr$.
Hence, coCEOs runs in $\BO B$ time.
Since we can govern the query time of coCEOs, we can use it on the budgeted MIPS setting~\cite{dWedge}.

\subsection{Make CEOs variants practical}

For CEOs variants, both indexing and querying requires Gaussian RP, which takes $\BO{dD}$ time for one point.
This cost is significant and often dominates the query time when $d$ is large.
Fortunately, there are several approaches to simulate the Gaussian RP.
We will use the Structured Spinners~\cite{Andoni15, Bojarski17,Choromanski17} that exploit the fast Hadamard transform to simulate Gaussian RP.

In particular, we generate 3 random diagonal matrices $\bD_1, \bD_2, \bD_3$ whose values are randomly chosen in $\{+1, -1\}$.
The Gaussian RP $\bR \bx$ can be simulated by the mapping $\bx \mapsto \bH \bD_3 \bH \bD_2 \bH \bD_1 \bx$ where $\bH$ is the Hadamard matrix.
With the fast Hadamard transform, the Gaussian RP can be simulated in $\BO{D \log D}$ time, which will not dominate the query time of CEOs variants.
%This approach has been used to speed up the hash computation~\cite{} in LSH framework.
Note that if $D$ is not a power of~2, we can simply add up additional zero coordinates.
More important, we can use universal hash functions to generate random diagonal matrices $\bD_i$.
Hence, it takes $\BO 1$ extra space for storing Gaussian RP simulation.
Furthermore, since both $\bH$ and $\bD_i$ are isometries, the inner products are preserved. 
Therefore, sCEOs-Est and sCEOs-TA can use these random rotated embeddings for computing inner products in post-processing, and hence do not need to keep the data set in the memory.

Figure~\ref{fig:CEO_Qs}~(b) shows that the Structured Spinners $\bH\bD_3\bH\bD_2\bH\bD_1$ can simulate well the Gaussian RP with almost the same MIPS precision but run $\BO{d / \log{D}}$ times faster.
Even though we use the Eigen library\footnote{\url{http://eigen.tuxfamily.org/index.php?title=Main_Page}} for the extremely fast C++ matrix-vector multiplication, Structured Spinners compute the signatures 4 times faster than the Gaussian RP on Nuswide with $d = 500$ and $D = 1024$.
Since our benchmark data sets are high-dimensional, and since we need $D$ is a power of~2, we set $D = 2^{\lfloor\log{d}\rfloor + 1} = \BO{d}$ in our experiments.

\subsection{Potential applications of CEOs variants}

In practice, we often need to run MIPS on large-scale data sets with a huge batch of queries in real-time at very high rates.
This section will discuss potential applications of CEOs variants on answering MIPS on distributed environments to handle such scenarios.

\textbf{sCEOs for distributed MIPS: }
Given $s$ concomitants and $D$ random projections, sCEOs answer query in $\BO{D\log D + db}$ time by computing $b$ inner products from the list $L_{IJ}$.
%Since we can store such list in one machine.
This key property makes sCEOs suitable for parallel computing.
\begin{enumerate}
	\item The size of $L_{IJ}$ is small enough to keep in memory for fast query response.
	\item Since we only assess one machine containing  $L_{IJ}$, sCEOs has almost no query broadcast overhead, and therefore yields large query throughput.
\end{enumerate}

\textbf{coCEOs for streaming MIPS: }
coCEOs uses $B = o(n)$ samples for computing partial inner products.
Such a small amount of information will significantly reduce the shuffling cost of the MapReduce implementations.
Furthermore, the cost of updating the data structure for one point is $\BO{D \log D + D \log m}$ due to the maintaining of $2D$ sorted lists $L_j$ and $S_j$ where $j \in [D]$.
Therefore, coCEOs is suitable for streaming MIPS over large-scale data sets where both queries and data arrive at very high rates.

\textbf{CEOs variants on P2P MIPS: }
In P2P system, we can horizontally partition the data sets into several partitions and each node will store a partial of our data sets.
In other scenario, different nodes can involve the search process by using their own data sets.
In both cases, CEOs variants work efficiently.
At the indexing phase, all nodes have to construct the data structure using the same random projections.
Exploiting the fast Hadamard transform, we only need to send $\BO{1}$ bits presenting for the matrix $\bD_i$ through the network.
Regarding privacy preservation, the query node can send $\BO{s \log D}$ bits corresponding to $s$ dimensions of extreme order statistics.
The query node will receive the largest estimates from all nodes and select the node with the largest estimate for the final MIPS evaluation.

\section{Experiment}

We implement CEOs variants~\footnote{\href{https://drive.google.com/file/d/1cALMtc8u2027rRXc4XR14n4wBSsbrBSD}{https://drive.google.com/file/d/1cALMtc8u2027rRXc4XR14n4wBSsbrBSD}} and other competitors in C++ using -O3 optimization and conduct experiments on a 2.80 GHz core i5-8400 32GB of RAM with single CPU.
%\footnotemark[\ref{dropbox}].
We present empirical evaluations on top-$k$ MIPS to verify our claims, including: 
\begin{enumerate}
	\item CEOs provides very high MIPS recall using a small number of dimensions $s$, which is consistent with the theory of concomitants of extreme order statistics.
	\item On inferior choice that uses the concomitants of $\osQone$, sublinear sCEOs outperforms the LSH bucket algorithm on indexing and querying  regarding both space and time.
	\item Both sCEOs-TA and coCEOs outperform competitive top-$k$ MIPS solvers on many real-world data sets.
	\item coCEOs outperforms sCEOs-TA when requiring high search recalls, e.g. 90\%.
\end{enumerate}

We use the speedup over the bruteforce search to measure the efficiency and $P@b$ to measure the search recall since we use  post-processing with a small budget of $b$ inner product computations.
We consider top-10 MIPS and hence when $b = 10$, we do not need post-processing.
The measurements are defined as follows:
\begin{align*}
\textrm{P@b} &= |\textrm{Retrieved top-10 } \cap \textrm{ True top-10}|/10 \, , \\
\textrm{Speedup} &= \textrm{Running time of bruteforce }/\textrm{ Running time of algorithm.}
\end{align*}
	
\begin{table}[!h]
	\centering
	\caption{Overview of the data sets}	\label{tb:datasets}
	\begin{tabular}{p{0.1cm}|p{0.8cm}|p{1cm}|p{0.7cm}|p{0.8cm}|p{0.5cm}|p{1.1cm}|p{0.8cm}}
		& Cifar10 & Nuswide & Yahoo & Msong & Gist &  Imagenet & Tiny5m \\
		\hline
		$d$ 	& 3072  		& 500 	& 300 & 420 	& 960 	 	& 150 		& 384 \\
		$n$ 	& 49K 		& 270K 	& 625K & 1M 	& 1M 	 	& 2.3M 		& 5M \\
		%    MF & \cite{Cremonesi10} & \cite{Cremonesi10} & \cite{Yu17} & \cite{Cremonesi10}, \cite{Yu17}
	\end{tabular}
\end{table}

\subsection{MIPS solvers and data sets}

We implement CEOs variants, including (1) \emph{1CEOs} and \emph{sCEOs} as sublinear MIPS solvers, (2) practical variants including \emph{sCEOs-Est}, \emph{sCEOs-TA}, and \emph{coCEOs} which have $\TilO{dn}$ index construction time and space.
We implement recent LSH-based schemes, including \emph{SimpleLSH}~\cite{SimpleLSH} and \emph{RangeLSH}~\cite{RangeLSH}, which exploit \emph{SimHash}~\cite{SimHash} for top-$10$ MIPS.
We also compare our solutions with \emph{dWedge}~\cite{dWedge}, a representative sampling approach for budgeted MIPS.
We implement the bruteforce search and LSH hash evaluations with the Eigen-3.3.4 library for the extremely fast C++ matrix-vector multiplication.
%\footnote{\url{http://eigen.tuxfamily.org/index.php?title=Main_Page}}  

We conduct experiments on standard real-world large-scale data sets, as shown in Table~\ref{tb:datasets}.
We randomly extract~1000 points (e.g. 1000~user vectors for Yahoo) to form the query set.
%\footnote{\url{ https://drive.google.com/drive/folders/1BHpiaii6Ur0rKSy5c9AFVwAhfLUQMXFE}}, including Netflix, Yahoo, and Gist.
%For the sake of comparison, we use the Netflix-200 ($n = 17,770; d = 200$) from~\cite{Yu17}, Netflix-300 ($n = 17,770; d = 300$) and Yahoo ($n = 624,961; d = 300$) from~\cite{Cremonesi10}, and Gist ($n = 1,000,000; d = 960$).
%We note that the two Netflix versions were generated by different matrix factorization tools.
%For Netflix and Yahoo, the item matrices are used as the data points.
%We randomly pick 1000 users from the user matrices to form the query sets.
%For Gist, we extracted randomly 1000 query points from the query set.
All randomized results are the average of 5 runs of the algorithms.

\subsection{Search recall of sCEOs}

This subsection shows the search recall of sCEOs on top-10 MIPS by implementing sCEOs-Est.
%We note that the reported search recalls are exactly the result of the sublinear sCEOs with $\BO{D^s}$ space.
We set $D = 2^{\lfloor\log{d}\rfloor + 1}$ to exploit the fast Structured Spinners. 
We observe that this setting satisfies $D \approx n^{1/2}$ on most of our high-dimensional data sets. 

\begin{figure} [!t]
 	\centering
 	%\raggedleft
 	\includegraphics[width=1\columnwidth]{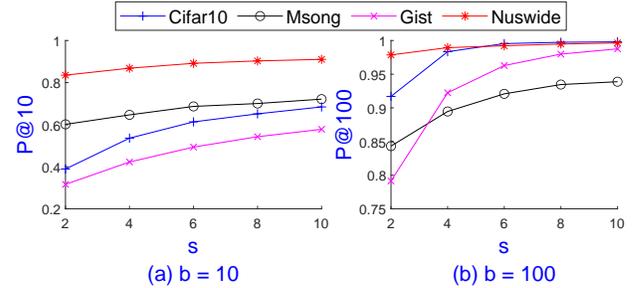}
 	\caption{Top-10 MIPS accuracy of sCEOs on several data sets with $D = 2^{\lfloor\log{d}\rfloor + 1}$, varying $s$: (a) $b = 10$ and (b) $b = 100$.}
 	\label{fig:sCEOs}
 \end{figure}

Figure~\ref{fig:sCEOs} presents the accuracy $P@b$ of top-10 MIPS when varying $s$ on Cifar10, Msong, Gist, and Nuswide.
It is clear that increasing $s$ leads to a substantial rise of the accuracy. 
When $s = 10, b = 100$, sCEOs achieves over 90\% accuracy on Msong and especially reaches nearly perfect recall on the other data sets.
The results confirm the reliability of our theoretical analysis for the sublinear sCEOs since the settings are similar to the requirements of Theorem~\ref{thm:main} where $D = n^{1/c}$ for any $c > 1$.

Given the above setting $D = \BO{d}, s = 10, b = 100$, sCEOs in practice uses $\BO{d^{10} + nd}$ space and $\BO{nd^{10}}$ time complexity for building the index and achieves nearly perfect recall with just 100 inner product computations on these data sets.

\subsection{Comparison on sublinear algorithms}

This subsection shows experiments comparing the performance between sublinear sCEOs and the $(l, L)$-parameterized LSH bucket algorithms with SimpleLSH and RangeLSH instances.
We note that both SimpleLSH and RangeLSH transform data and query into unit sphere in order to exploit SimHash.

While sCEOs can simply choose top-$b$ points with the largest inner product estimates in the list $L_{IJ}$, it is impossible to tune LSH parameters to return the best $b$ candidates for each query.
This is due to the fact that we do not know the collision probability between the query and data points.
Given a fixed number of hash tables $L$, the number of concatenating hash functions $l$ will control the number of collisions and therefore the candidate set size. 
%In particular, large values of $l$ will decrease the collision probability and hence candidate set size and vice versa.
The query complexity of LSH schemes are $\BO{dlL + db}$ where the first term is from the hash computation and the second term comes from the post-processing phase.

\begin{figure} [!t]
	\centering
	%\raggedleft
	\includegraphics[width=1\columnwidth]{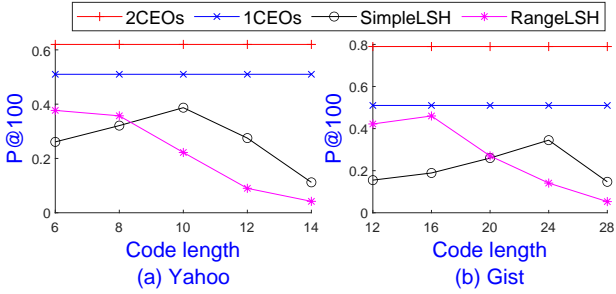}
	\caption{Comparison of top-10 MIPS accuracy of 1CEOs, 2CEOs, SimpleLSH and RangeLSH with $b = 100$ and $D = 2^{\lfloor\log{d}\rfloor + 1}$ on (a) Yahoo and (b) Gist.}
	\label{fig:1CEOs}
\end{figure}

For LSH parameter settings, we first fix $L = 512$ as suggested in previous LSH-based MIPS solvers~\cite{SimpleLSH,Shrivastava14} and vary the parameter $l$ such that we achieve the highest search recall given $b$ candidates for each query.
RangeLSH uses $p = 4$ partitions and each partition has $L/p$ hash tables.
We observe that increasing $p$ will decrease the performance.
%We report a wide range of values $l$ with the highest search recall.
We use the Eigen library to speed up the hash computation.
Regarding sCEOs, we only implement the inferior choices: \emph{1CEOs} with $s = 1$ and \emph{2CEOs} with $s = 2$ where we use concomitants of $\osQone$ and $\osQd$.
Due to the similar results, Figure~\ref{fig:1CEOs} shows the representative results on Yahoo and Gist.

It is clear that CEOs schemes are superior to LSH schemes regarding search recall on these two data sets.
We observe that RangeLSH reaches the highest search recall with smaller $l$ than SimpleLSH.
Then, the accuracy of both schemes decreases when increasing $l$ on both data sets.
A simple computation indicates that the average top-1 MIPS value after the SimpleLSH transformation is 0.24 on Yahoo.
Hence, the total number of collisions is at most $nL\tau_1^h \approx 1$ when $L = 512$, $h = 14$.

We observe that the average top-10 MIPS values are almost same even with $1000 > L$ partitions for RangeLSH.
Hence there might be no advantages of increasing the distance gap between ``close'' and ``far way''  points.
Since the top-$k$ MIPS points are often distributed on the same partition due to the similar large 2-norm values, RangeLSH reaches the highest search recall with smaller $l$ than SimpleLSH.
Since each partition will have $L/4$ hash tables, increasing $l$ will decrease the accuracy of RangeLSH faster than SimpleLSH, as observed in Figure~\ref{fig:1CEOs}.

While 2CEOs gives higher search recalls than 1CEOs, its time and space complexity of building the index are also more significant.
Table~\ref{tab:1CEOs} shows a comparison of 1CEOs, 2CEOs, SimpleLSH~24 bits and RangeLSH~16 bits on indexing and querying on Gist.

\begin{table}[!h]
	\centering	
	\caption{Comparison of indexing and querying between 1CEOs, 2CEOs, SimpleLSH ($l = 24$) and RangeLSH ($l = 16, p = 4$) on Gist when fixing $b = 100$ and $L = 512$.}
	\label{tab:1CEOs}
	\begin{tabular}{|c|c|c|c|c|} 
		\hline
		\multirow{2}{*}{Algorithms} & \multicolumn{2}{c|}{Index} & \multicolumn{2}{c|}{Query} \\ 
		\cline{2-5} 
		& Time & Space & $P@100$ & Speedup \\ 
		\cline{2-5}
		\hline
		1CEOs & \textbf{1.5 mins} & 0.8GB & 51\% & \textbf{198}$\times$ \\ 
		\hline
		2CEOs & 1.2 hours & 8.4GB & \textbf{80\%} & \textbf{145}$\times$ \\ 
		\hline 
		SimpleLSH\textsubscript{24} & 1.5 hours & 10GB & 35\% & 62$\times$ \\ 		 
		\hline
		RangeLSH\textsubscript{16} 	& 1.1 hours & 10GB & 46\% & 52$\times$ \\ 
		\hline
		%		RP\textsubscript{10} & 5s & 7.7GB & 50\% & 34$\times$ \\ 
		%		\hline		
	\end{tabular}
\end{table}

\begin{figure*} [!t]
	\centering
	%\raggedleft
	\includegraphics[width=1\textwidth]{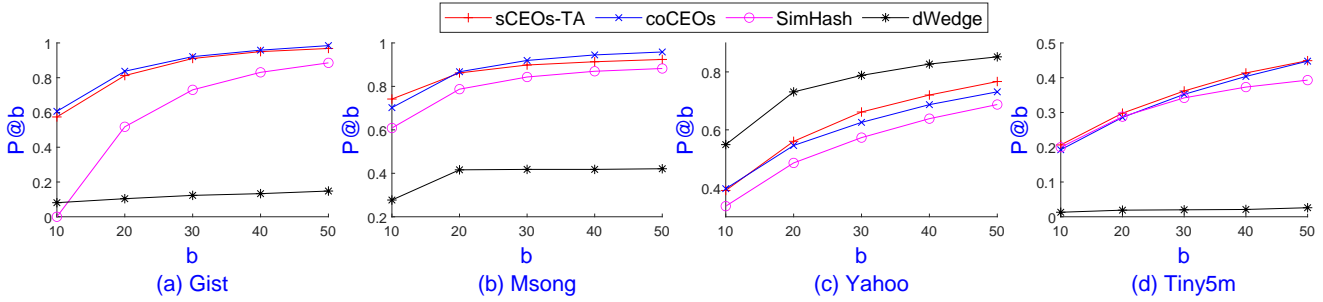}
	\caption{Comparison of top-10 MIPS accuracy between sCEOs-TA, coCEOs, SimHash, and dWedge with $D = 2^{\lfloor\log{d}\rfloor + 1}$, $s = 10$ for sCEOs-TA and $s' = 20$ for coCEOs, and varying $b$ on: (a) Gist, (b) Msong, (c) Yahoo, and (d) Tiny5m.}
	\label{fig:sCEOs-TA}
\end{figure*}

It is clear that 1CEOs and 2CEOs outperform LSH schemes regarding both accuracy and efficiency on indexing and querying. 
Regarding the index, since the data set itself requires~8GB, 2CEOs uses only 0.4GB extra storage compared to 2GB of LSH schemes.
Because 1CEOs only keeps $b = 100$ points on each of $D = 1024$ dimensions, its index storage is just 10\% of the data.
Furthermore, the index construction time of 1CEOs is just some minutes compared to more than~1 hour of LSH schemes and 2CEOs.

Regarding querying, we observe that the hash computation time dominates the total time of LSH schemes with more than 70\%, whereas more than 97\% of total time of 1CEOs and 2CEOs are used for computing $b = 100$ inner products.
While 1CEOs provides lower search recall than 2CEOs, ones can simply increase $b$ inner product computations in post-processing to boost the accuracy.
We observe that 1CEOs with $b = 500$ achieves almost the same search recall of 2CEOs with just 40$\times$ speedup.
Overall, 1CEOs is superior to LSH schemes regarding the indexing and querying.

We observe that the accuracy of 2CEOs is consistent with sCEOs-Est with $s = 2$, as shown in Figure~\ref{fig:sCEOs}~(b). 
In other words, ones can foresee the performance of sCEOs by implementing and testing sCEOs-Est in a few minutes.
For example, sublinear sCEOs can achieve the performance of at least 85\% accuracy for $s = 2$ and at least 90\% for $s = 4$ with approximately 150$\times$ speedup on Cifar10, Msong and Nuswide.
In contrast, we spent a day to find the best parameter settings for LSH schemes.

It is worth emphasizing that almost 100\% of the search time of sCEOs are dedicated for computing inner products. 
Specifically, sCEOs runs on a single CPU and answers~1000 queries just only in~2 seconds on Gist.
Ones can simply scale up the search process with multiple CPUs or distributed machines since sCEOs does not require any communication between CPUs or machines.

\subsection{Comparison on estimation algorithms}

This subsection shows experiments comparing the performance between CEOs variants, including sCEOs-Est, sCEOs-TA, coCEOs, dWedge, and LSH code estimation schemes, including SimHash, SimpleLSH and RangeLSH on Gist, Msong, Yahoo, and Tiny5m.
We note that Gaussian RP often suffers lower accuracy than these algorithms so we do not report it here.

Regarding parameter settings, sCEOs-Est and sCEOs-TA use $D = 2^{\lfloor\log{d}\rfloor + 1}$, $s = 10$ and they share the same accuracy.
LSH schemes use the code length $l = 128$ and use the \texttt{\_builtin\_popcount} function of compilers for Hamming distance computation.
We observe that LSH with $l = 128$ shares a similar speedup with sCEOs-Est.

While sCEOs-Est and LSH schemes require $\BO{n}$ times for answering MIPS, sCEOs-TA often runs much faster.
This is because sCEOs-TA might not have to estimate $n$ inner products.
To show the sublinear running time, we will compare sCEOs-TA with coCEOs and dWedge using $B = n/100$.
dWedge shares a similar speedup with coCEOs due to the same mechanism: using $B$ samples to compute partial estimates of inner products.
Note that we set $s' = 20$ for coCEOs since it can use a larger number of extreme order statistics.

Figure~\ref{fig:sCEOs-TA} shows the accuracy $P@b$ for a wide range of $b$ inner products in post-processing of sCEOs-TA, coCEOs, SimHash, dWedge on Gist, Msong, Yahoo, and Tiny5m.
It is clear that CEOs variants consistently outperform SimHash on all 4 data sets while dWedge only shows advantages on the recommender system Yahoo.
With $b = 50$, both sCEOs-TA and coCEOs achieve at least 90\% accuracy on Gist and Msong whereas SimHash gives just above 80\%.
CEOs variants can provide 45\% accuracy on Tiny5m whereas SimHash achieves approximately 35\%.
On Tiny5m, we can set $b = 100$ and $s = 20$ to increase $P@b$ of sCEOs-TA to at least 80\% while SimHash still suffers from very low accuracy.
%The result again confirms the reliability of Theorem~\ref{thm:main} that small values of $s$ and $b$ suffice for top-$k$ MIPS.

Table~\ref{tb:sCEOsTA} shows a detailed comparison of three CEOs variants and three LSH variants on indexing and querying with $b = 50$ on Gist.
%The result is also similar on other data sets.
SimHash provides the highest search recall and similar speedup among LSH schemes given the same code length.
It is reasonable since SimHash uses the norms directly to estimate inner products.
While SimHash 256 bits and CEOs variants achieve similarly almost perfect recall, their speedups are very different.
SimHash 256 bits are~3 times slower than sCEOs-Est and~15 times slower than sCEOs-TA and coCEOs.
We observe that sCEOs-Est with $s = 10$ shares the same speedup with LSH schemes of $l = 64$ bits due to the fast addition operators supported by the Eigen library.

We note that the average inner products of top-10 MIPS on Gist is~0.68 and it does not change even with~1000 partitions on RangeLSH.
Their Hamming distances are approximately 14 and 28 for 64 bits and 128 bits, respectively.
This means that if you use a single hash table with multi-probing, you have to probe several thousands points for inner product computations, as illustrated in the experiment of RangeLSH.
Such a large number of inner products will degrade the search performance, especially when data are located on disk.
On the other hands, SimHash and CEOs variants only compute $b = 50$ inner products and achieve more than 90\% accuracy.

Though sCEOs-TA uses more time and space to construct the index, sCEOs-TA and coCEOs achieve almost the same search recall and speedup.
Empirically, we observe that sCEOs-TA needs a few thousands of inner product estimates on all 4 data sets.
This cost is nearly the same cost of computing partial inner products with $B = N/100$ of coCEOs and dWedge.
Since we use the same $b$ in post-processing, coCEOs and sCEOs-TA achieve similar speedup on these data sets.
For example, sCEOs-TA and coCEOs achieve more than 90\% accuracy with at least 150$\times$ speedup on Msong and Nuswide, but we do not report in details here.

\begin{table}[!t]
	\centering
	\caption{Comparison of indexing and querying with $b = 50$ in post-processing between sCEOs-Est, sCEOs-TA, coCEOs, SimHash, SimpleLSH, RangeLSH ($p = 4$) on Gist.}
	\label{tb:sCEOsTA}
	\begin{tabular}{|c|c|c|c|c|} 
		\hline
		\multirow{2}{*}{Algorithms} & \multicolumn{2}{c|}{Index} & \multicolumn{2}{c|}{Query} \\ 
		\cline{2-5} 
		& Time & Space & $P@50$ & Speedup \\ 
		\cline{2-5}
		\hline
		sCEOs-Est & 74s & 8.2GB & \textbf{97\%} & 45$\times$ \\ 
		\hline
		sCEOs-TA & 307s & 16.4GB & \textbf{97\%} & \textbf{166}$\times$ \\ 
		\hline 
		coCEOs 	& 172s & 8GB & \textbf{98\%} & \textbf{194}$\times$ \\
		\hline
		SimHash\textsubscript{256} & 60s & 8GB & \textbf{97\%} & 12$\times$ \\ 		 
		\hline
		SimHash\textsubscript{128} & 30s & 8GB & 89\% & 25$\times$ \\ 
		\hline
		SimpleLSH\textsubscript{128} 	& 30s & 8GB & 73\% & 37$\times$ \\ 
		\hline
		RangeLSH\textsubscript{128} 	& 53s & 8GB & 76\% & 27$\times$ \\ 
		\hline
		%		RP\textsubscript{10} & 5s & 7.7GB & 50\% & 34$\times$ \\ 
		%		\hline		
	\end{tabular}
\end{table}

Since sCEOs-TA extracts top-$b$ maximum inner product estimates, sCEOs-TA can be used to speed up the index construction of sublinear sCEOs.
In particular, ones can run in parallel $\BO{(D/s)^s}$ sCEOs-TA instances to construct the sCEOs index.

\subsection{coCEOs vs. sCEOs-TA on high search recalls}

In the previous subsection, we observe that sCEOs-TA with $s = 10$ achieves similar search recall and speedup with coCEOs with $s' = 20$ on all 4 data sets.
However, to achieve significantly high search recalls, sCEOs-TA needs to use larger values of $s$ and $b$.
For example, Tiny5m requires $b = 100$, $s = 40$ to achieve 90\% recall.
On these settings, sCEOs-TA's performance is deteriorated and even outperformed by sCEOs-Est.
This subsection demonstrates these findings on Yahoo, Tiny5m, and Imagenet, and shows that coCEOs is often superior to sCEOs-TA for a wide range of $s$ and $b$.

\begin{figure} [!t]
	\centering
	%\raggedleft
	\includegraphics[width=1\columnwidth]{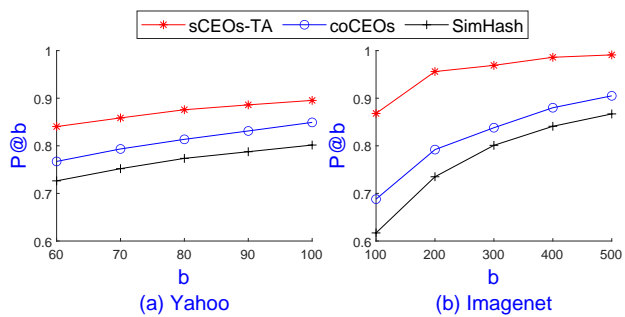}
	\caption{Top-10 MIPS accuracy of sCEOs-TA, coCEOs, and SimHash with varying $b$ on (a) Yahoo and (b) Imagenet.}
	\label{fig:coCEOs}
\end{figure}

We observe that sCEOs-TA requires $s = 20$, $b = 100$ and $s = 100$, $b = 500$ to achieve 90\% accuracy on Yahoo and Imagenet, respectively.
Since SimpleLSH and RangeLSH are outperformed by SimHash, we present only SimHash's results with $l = 128$ and $l = 512$ bits code on Yahoo and Imagenet, respectively.
coCEOs uses $B = n/100$, $s' = 2s = 40$ on Yahoo and $B = n$, $s' = s = 100$ on Imagenet.
All methods use the same $b$ inner product computations in post-processing.
Figure~\ref{fig:coCEOs} shows a comparison of coCEOs, sCEOs-TA, and SimHash on Yahoo and Imagenet.

While coCEOs outperforms SimHash, providing 85\% and 90\% search recall on Yahoo and Imagenet, respectively, sCEOs-TA achieves the highest recall with at least 90\% on both data sets.
However, the running time of sCEOs-TA is deteriorated due to the large values of $b$ and $s$.
Table~\ref{tb:coCEOsYahoo} shows a detailed comparison of these methods on the querying process on Yahoo.

\begin{table}[!h]
	\centering
	\caption{Comparison of querying performance with $b = 100$ in post-processing between sCEOs-Est, sCEOs-TA, coCEOs, and SimHash ($l = 128$) on Yahoo.}
	\label{tb:coCEOsYahoo}
	\begin{tabular}{|c|c|c|c|c|} 
		\hline		
		Algorithm & SimHash\textsubscript{128} & sCEOs-Est & sCEOs-TA & coCEOs  \\ 
		\hline
		Speedup & $7\times$ & $9\times$ & $10\times$ & \textbf{116}$\times$ \\ 
		\hline
		$P@100$ & 80\% & 90\% & 90\% & 85\% \\ 
		\hline 
	\end{tabular}
\end{table}

It is clear that coCEOs runs orders of magnitude faster than sCEOs-TA, sCEOs-Est and SimHash 128 bits.
We note that coCEOs with $b = 150$ can boost the accuracy up to 90\% and still offer at least 100$\times$ speedup.

On Imagenet, we observe that both sCEOs-TA and sCEOs-Est only provide marginal speedup due to the large reduced dimensions, i.e. $s = 100$ and $d = 128$.
However, coCEOs provides superior speedup to both sCEOs-TA and SimHash 512 bits while maintaining 90\% accuracy since it uses $B = n$.
It achieves 7$\times$ speedup and 90\% accuracy whereas SimHash 512 bits gives only 2$\times$ speedup and 87\% accuracy with $b = 500$.
We note that since all methods use the same $b$ inner product computations in post-processing, decreasing $b$ can significantly increase the speedup with the trade of accuracy.

On Tiny5m, sCEOs-TA also shows a slow query time with $b = 100$, $s = 40$.
coCEOs with $b = 500$, $s' = 40$ can achieve 90\% accuracy with at least 100$\times$ speedup whereas SimHash 128 bits gives only above 50\% accuracy with just 16$\times$ speedup.

\section{Related Work}

Due to the ``curse of dimensionality'', there is no known algorithms for solving exactly MIPS in truly sub-linear time.
In fact, if there exists such an algorithm, the Strong Exponential Time Hypothesis (SETH)~\cite{SETH}, a fundamental conjecture in computational complexity theory, is wrong~\cite{Rubinstein18,Williams14}.
Therefore, sequential scanning with pruning the search space techniques has been used to speed up MIPS and to return exact results~\cite{FEXIPRO,LEMP}.
However, these solutions run in $\Theta(n)$ time.
In contrast, this work investigates algorithms that run in $o(n)$ time and provide approximate top-$k$ MIPS.
Our empirical evaluation on standard large-scale data sets shows that it is possible to achieve 90\% search recall given $o(n)$ query time.

An alternative efficient solution is applying sampling methods to estimate the vector-matrix multiplication derived by top-$k$ MIPS~\cite{Diamond,Wedge}.
The basic idea is to sample a point $\bx$ with probability proportional to the inner product $\dotxq$.
The larger inner product values the point $\bx$ has, the more occurrences of $\bx$ in the sample set.
However, the number of required samples are often much larger than $n$ to guarantee high quality approximate MIPS.

It is worth noting that our investigated problem is similar to the budgeted MIPS, which has been recently studied in~\cite{dWedge,Yu17}.
These solutions also require to construct the index in $\TilO{dn}$ time and space.
We have demonstrated that our proposed solutions achieve significantly higher accuracy on many real-world data sets.

Recently, alternates to data-independent LSH are data-dependent schemes for approximate MIPS, including product quantization~\cite{Faiss,Guo16} and similarity graphs~\cite{Malkov20}. 
Since these methods lack rigorous theoretical guarantees, we do not know when these schemes work or fail. 
Furthermore, these methods require significant data-dependent indexing time, and hence cannot be used on many applications, e.g. streaming search, scaling up machine learning models where both data and query distributions can change. 
In particular, quantization indexing runs as slowly as a $k$-mean clustering and might converge to local minimum which degrades the performance. 
Building an exact Delauney graph is unfeasible due to the exponentially growing number of edges in high dimension~\cite{Malkov20}. 
A popular approach is to build an approximation of Delaunay graph~\cite{Morozov18,Zhou19}, which still requires $\Osymbol(n^2)$ time and does not offer any theoretical guarantee on search performance.
In our work, we choose to compare with LSH~\cite{SimpleLSH,RangeLSH} due to the data-independent scheme with theoretical guarantees on query performance, and sub-quadratic (or linear) indexing time and space.

We note that CEOs shares some similar spirit with BOND~\cite{Vries02}, a branch-and-bound search approach that selects a few of important data dimensions for execution.
Since BOND is executed on the original data space, there is no theoretical guarantee on the search performance and it might run as slow as the sequential scanning.
On the technical side, CEOs uses the maximum of Gaussian random variables for inner product estimation. 
This idea is similar to~\cite{Hadar19A,Hadar19B} which study the maximum of Gaussian variables for estimating the correlation of two parties' variables in the distributed setting given a minimum number of exchanged bits.
However, CEOs exploits the theory of concomitants of extreme order statistics to guarantee the search recall and optimality with the elementary Chernoff bounds.
%By the end of the sampling process, we retrieve top-$m$ budgeted points ($m > k$) with largest occurrences in the sample set via a counting histogram.
%The top-$k$
%points, with the largest inner product with the query
%%largest inner product points from
%among these $m$ points,
%%with the query
%will be returned as an approximation for top-$k$ MIPS.
%It is clear that sampling schemes naturally fit to the budgeted setting since the more samples we use, the higher accuracy we can achieve.
%However, the linear cost of scanning \emph{all} data points to return top-$m$ candidate points limits sampling methods to $o(n)$ budget. 

%\begin{figure} [!t]
%	\centering
%	%\raggedleft
%	\includegraphics[width=1\columnwidth]{Fig/coCEOs_Param.eps}
%	\caption{Accuracy of top-10 MIPS with varying $s_0$ and $b = 100$ : (a) Yahoo with $D = 512$ and (b) Mnist with $D = 1024$.}
%	\label{fig:coCEOs}
%\end{figure}

\section{Conclusions}\label{sec:conclusion}

The paper proposes CEOs, a novel dimension reduction for top-$k$ MIPS based on the theory of concomitants of extreme order statistics.
Utilizing the asymptotic behavior of these concomitants, we show that CEOs provides a sublinear query time with an exponential space complexity.
The search recall can be theoretically guaranteed under a mild condition of data and query distributions.

To deal with the exponential space and time complexity of indexing, we propose two variants, including sCEOs-TA and coCEOs, using linear space to solve top-$k$ MIPS.
While sCEOs-TA outperforms competitive MIPS solvers regarding both efficiency and accuracy, coCEOs is even superior over a wide range of parameter settings.
Empirically, they achieve more than 100$\times$ speedup compared to the bruteforce search while returning top-10 MIPS with accuracy at least 90\% on many real-world large-scale data sets.
%\end{document}  % This is where a 'short' article might terminate

%ACKNOWLEDGMENTS are optional

\section{Acknowledgments}
We thank Rasmus Pagh for pointing to the similar technique used in distributed statistical inference under communication constraints~\cite{Hadar19B}.


%\vspace{-1mm}

\bibliographystyle{abbrv}
\balance
\bibliography{sigproc}

  % sigproc.bib is the name of the Bibliography in this case
% ACM needs 'a single self-contained file'!
%\balancecolumns

%\balancecolumns % GM June 2007
% That's all folks!
\end{document}